\documentclass[12pt,showpacs,twocolumn,preprintnumbers,aps,prr,reprint,superscriptaddress]{revtex4-1}

\usepackage{graphicx}
\usepackage{subfigure}
\usepackage[labelfont={footnotesize,bf},textfont={footnotesize}]{caption}
\usepackage{color}
\usepackage{amsmath,amssymb,mathtools,xfrac}
\usepackage{epstopdf}
\usepackage[linktocpage,colorlinks=true,linkcolor=blue,citecolor=blue, urlcolor = black, breaklinks=true]{hyperref}
\usepackage{url}
\usepackage{verbatim}
\usepackage{breakcites}
\usepackage[version=3]{mhchem}

\newcommand{\be}{\begin{equation}}
\newcommand{\ee}{\end{equation}}

\DeclareMathOperator{\Var}{Var}

\newcommand{\kB}{k_{\mathrm{B}}}

\makeatother

\begin{document}


\title{The Thermodynamic Cost of Erasing Information in Finite-time}

\author{L. T. Giorgini}
\thanks{These two authors contributed equally}
\email{ludovico.giorgini@su.se}
\affiliation{Nordita, Royal Institute of Technology and Stockholm University, Stockholm 106 91, Sweden}

\author{R. Eichhorn}
\thanks{These two authors contributed equally}
\email{eichhorn@nordita.org}
\affiliation{Nordita, Royal Institute of Technology and Stockholm University, Stockholm 106 91, Sweden}

\author{M. Das}
\email{moupriya@iitmandi.ac.in}
\affiliation{Indian Institute of Technology Mandi, Kamand, Himachal Pradesh 175075, India}

\author{W. Moon}
\email{woosok.moon@gmail.com}
\affiliation{Department of Environmental Atmospheric Sciences, Pukyong National University, 48513 Pusan, South Korea}

\author{J. S. Wettlaufer}
\email{john.wettlaufer@yale.edu; jw@fysik.su.se}
\affiliation{Yale University, New Haven, Connecticut 06520, USA}
\affiliation{Nordita, Royal Institute of Technology and Stockholm University, Stockholm 106 91, Sweden}
\date{\today}
\begin{abstract}
The Landauer principle sets a fundamental thermodynamic constraint on the minimum amount of heat that must be dissipated to erase one logical bit of information through a quasi-statically slow protocol.
For finite time information erasure, the thermodynamic costs depend on the specific physical realization of the logical memory and how the information is erased.  Here we treat the problem within the paradigm of a Brownian particle in a symmetric double-well potential. The two minima represent the two values of a logical bit, 0 and 1, and the particle's position is the current state of the memory.  The erasure protocol is realized by applying an external time-dependent tilting force.  We derive analytical tools to evaluate
the work required to erase a classical bit of information in finite time via an arbitrary continuous erasure protocol, which is a relevant setting for practical applications. Importantly, our method is not restricted to the average work, but instead gives access to the full work distribution arising from many independent realizations of the erasure process. Using the common example of an erasure protocol that changes linearly with time \textcolor{black}{acting on a double-parabolic potential}, we explicitly calculate all
relevant quantities and verify them numerically.
\end{abstract}

\maketitle

\section{Introduction}
\label{sec:intro}

A central goal in information technology is to improve the speed of computational processes.  A problematic consequence is the substantial unavoidable generation of heat, with deleterious consequences for the devices themselves and the environment at large. Moreover, improved functionality and more complex computational tasks involve consumption of a larger amount of electrical energy, which is eventually dissipated  in the computational devices themselves~\cite{chiu}.

Beyond Ohmic dissipation in the electronic elements of the computational device, there is an additional heat production related to the irreversible logical steps that constitute the essential ``building blocks'' of computational tasks~\cite{gregg}.  Indeed, fundamental thermodynamic principles link the manipulation of bits of information to heat dissipation associated with an increase in the entropy of the environment.  This linkeage  is known as ``Landauer's principle'', which states that erasing one logical bit of information dissipates at least an amount $\kB T \ln 2$ of heat \cite{Landauer1961,land1988}, where $\kB$ is Boltzmann's constant and $T$ is the temperature of the environment. The Landauer bound is only achieved when the erasure process is {\em quasi-statically slow,} whereas in any faster, or finite-time, process the heat dissipated exceeds $\kB T \ln 2$.
In addition to logical computation, the Landauer principle is relevant in the measurement and storage of information, because erasure is required for ``reset to zero'' operations~\cite{land1988,parr}.

Thermal noise plays a noticeable role when the memory states representing a bit of information are implemented in a ``small'' mesoscopic system consisting of only a few degrees of freedom with characteristic energies of the order of the thermal energy $\kB T$. Thus, the dissipated heat becomes a fluctuating quantity \cite{lutz} and the Landauer bound refers to an average over many realizations of an erasure process.  In the context of stochastic thermodynamics in such mesoscopic systems \cite{jarzynski,seifert,broeck1,seifert1},  the thermodynamic and information-theoretic implications of the Landauer principle are a very active area of research
\cite{land1988,piech,lutz,saga,toya,broeck,berut,berut1,jun,rold,zulk,zulk1,espo,boyd,boyd1,bech,bech1,aure1,desh,dago,dago1,lee2022}.

It is clear that for applications, rather than quasi-statically slow erasure processes, it is essential to quantify the dissipated heat when erasure occurs in finite time~\cite{aure1,espo,zulk,zulk1,bech,bech1,dago,dago1,lee2022}.  Thus a key goal is to find optimal protocols such that a bit of information is erased quickly, reliably and with the minimal dissipation of heat \cite{aure1,zulk,zulk1,boyd1,bech,bech1} (see also \cite{aure}).

We can treat one-bit memory erasure physically as a Brownian particle moving in a double-well potential \cite{land1991,lutz,zulk,berut,berut1,jun,bech,bech1,desh,dago,dago1}.
The two potential minima represent the two values of the information bit, and the particle position represents the current state of the memory, which can be manipulated by applying external forces and executing the
 ``erasure protocol''.  Despite the apparent simplicity of this model, theoretical results of the distribution of dissipated heat, or  the work required to perform the erasure procedure, are limited.  Indeed,  as far as we are aware,  for such classical bi-stable systems the only theoretical predictions beyond the Landauer bound produce bounds for the \emph{average} work that is required to erase one bit of information in finite time \cite{aure1,zulk,bech,bech1}.  Importantly, the minimum average work expended under optimal conditions is inversely proportional to the protocol duration, with a system-- and protocol--specific proportionality factor \cite{aure1,zulk,zulk1,boyd1,bech,bech1}.  This finding has been reproduced experimentally \cite{berut,berut1,jun,dago,dago1}.

Firstly, in \S \ref{sec:systemetc}, we describe our model for a Brownian particle in a driven double-well potential and define our central observable; the work required to erase one logical bit of information.
Secondly, we detail an analytical method of calculating the average work, and the higher order moments of the complete work distribution, for fast erasure protocols with arbitrary (continuous) time dependence (\S \ref{sec:work}).  Our main results are explicit formulae  for the average work \eqref{eq:<W>} and its variance \eqref{eq:var(W)}.
\textcolor{black}{%
They are expressed in terms of the statistics of the jump times between the two potential wells that represent the
memory states, and in terms of the shape of these potential wells encoded in a partition function (understood as a sum over time-dependent, locally equilibrated states)}.
We find that the principal sources of randomness are the transitions over the potential barrier between the two memory states, whereas the fluctuations within the potential wells have negligible effects on the work distribution.  Thirdly,
\textcolor{black}{%
we provide a concrete recipe for calculating the jump-time distributions \eqref{eq:P(t_i)} (or \eqref{eq:P(t_i)givenPIn}) that can be applied to a wide range of potentials and protocols.
}

By combining these results, the averages in \eqref{eq:<W>} and \eqref{eq:var(W)} reduce to simple numerical quadratures,
\textcolor{black}{%
which only require the shape of the unperturbed double-well potential and the time change of the applied erasure protocol as ingredients.
Our theoretical approach relies on an essential, but not very restrictive, assumption: \emph{relaxation within the potential wells} represents the {\em fastest deterministic} time scale in the system.  This imposes an upper bound on the rate of change of the erasure protocol. 
In fact, this assumption is obeyed extremely well in standard experiments on information erasure \cite{berut,berut1,jun},
such that all our general analytical results are directly applicable \cite{note0}.
}

\textcolor{black}{%
Finally, as specific example, we apply our general theory to a double-well potential constructed of two harmonic traps and an erasure protocol, which changes linearly in time (\S \ref{sec:F(t).lin}). We} calculate all relevant quantities and compare our analytical predictions to numerical simulations.

\section{The system, the dynamics and the central observable}
\label{sec:systemetc}
\subsection{Model structure}
\label{sec:model}

We consider the Brownian motion of a particle with position $X(t)$ within a double-well potential $U(X,t)$,
\textcolor{black}{%
with the unperturbed potential $U(X)=U(X,t=0)$ having minima at $X = \pm a$ and a relative maximum at $X=0$.
We construct $U(X)$ from a single-well potential $V(X)$ with a unique minimum at $X=0$
and with $V(X \to \pm\infty) \to \infty$,
\begin{equation}
\label{eq:Ugeneral}
U(X)=
\begin{cases}
\begin{array}{l@{\hspace{4ex}}l}
V(X+a)=V(a+X)      & \textrm{for } X < 0 \\
V(-(X-a)) = V(a-X) & \textrm{for } X \geq 0
\end{array}
\end{cases}
\, .
\end{equation}
This construction will prove convenient when calculating the moments of the work distribution; the condition $V(X \to \pm\infty) \to \infty$ is needed to ensure the existence of a stationary distribution within $V(X)$, but the details of how $V(X)$ approaches infinity, in particular beyond the point at which $V(a+X)$ and $V(a-X)$ are joined, are irrelevant. 
Note that Eq.~\eqref{eq:Ugeneral} covers any mirror-symmetric double-well potential, as, e.g., the quartic double well with a smooth maximum between the two wells (choose $V(X)=K X^2(X-2a)^2$ for $X \leq a$ with some $K>0$ and an arbitrary monotonically increasing function ``attached to'' $V(a)=K a^4$ for $X > a$), or the double-parabolic potential specified below in Eq.~\eqref{eq:U} with a cusp-shaped maximum (choose $V(X)=(K/2)X^2$).
}

The memory states 0 and 1 refer to the particle being located in the left and right well of the potential respectively, thereby characterizing the storage of \textcolor{black}{one bit} of information \cite{parr}.
We assume that initially ($t=0$) the particle is in equilibrium with an environment at temperature $T$, such that both states 0 and 1 are occupied with
equal probability $1/2$.  The stability of the memory states requires the potential barrier between the two wells to be much larger than the thermal energy \cite{parr}.

A standard protocol for erasing such information bits is to force both states into the same final state \cite{parr}.  This can be achieved by applying an external forcing protocol $F(t)$, which at the end of the erasure process must be ``switched off'' or ``reset'' so that the double-well potential returns to its initial, unperturbed configuration.  Therefore, the forcing protocol of total duration $\tau$ consists of an ``erasure phase'' $F_{\mathrm{erase}}(t)$
for $0 \leq t \leq \tau^*$ (with $\tau^* < \tau$), and a rapid ``resetting phase'' $F_{\mathrm{reset}}(t)$ within a short time interval $\tau-\tau^* \ll \tau^*$, such that
\begin{subequations}
\label{eq:protocol}
\begin{equation}
\label{eq:F(t)}
F(t) = \left\{
\begin{array}{ll}
F_{\mathrm{erase}}(t) & \mbox{for } 0 \leq t \leq \tau^* \\
F_{\mathrm{reset}}(t) & \mbox{for } \tau^* < t \leq \tau
\end{array}
\right.
\, ,
\end{equation}
which fulfills the two constraints;
\begin{gather}
\label{eq:F(t=0)}
F_{\mathrm{erase}}(0) = F_{\mathrm{reset}}(\tau) = 0 \, ,
\\
\label{eq:F(t=tau*)}
F_{\mathrm{erase}}(\tau^*) = F_{\mathrm{reset}}(\tau^*) =
\textcolor{black}{\max_{-a \leq X \leq 0} \left[ U'(X) \right]
}
\, ,
\end{gather}
\end{subequations}
\textcolor{black}{with the prime denoting differentiation with respect to $X$.}
The condition \eqref{eq:F(t=tau*)} is synonymous with choosing  the right well (state 1) as the final state of the erasure phase.   For a tilting force of $F_{\mathrm{erase}}(\tau^*) = \textcolor{black}{\max_{-a \leq X \leq 0} \left[ U'(X) \right]}$ the left potential well (state 0)
disappears completely, leaving the right well (state 1) as the only minimum of the potential, and thus the particle ends up in state 1 independent of its initial state.
During the time interval $(\tau^*,\tau]$ the potential is rapidly brought back to its initial configuration, during which the particle still resides in state 1.

\begin{figure}
\begin{center}
\includegraphics[width=1.0\linewidth]{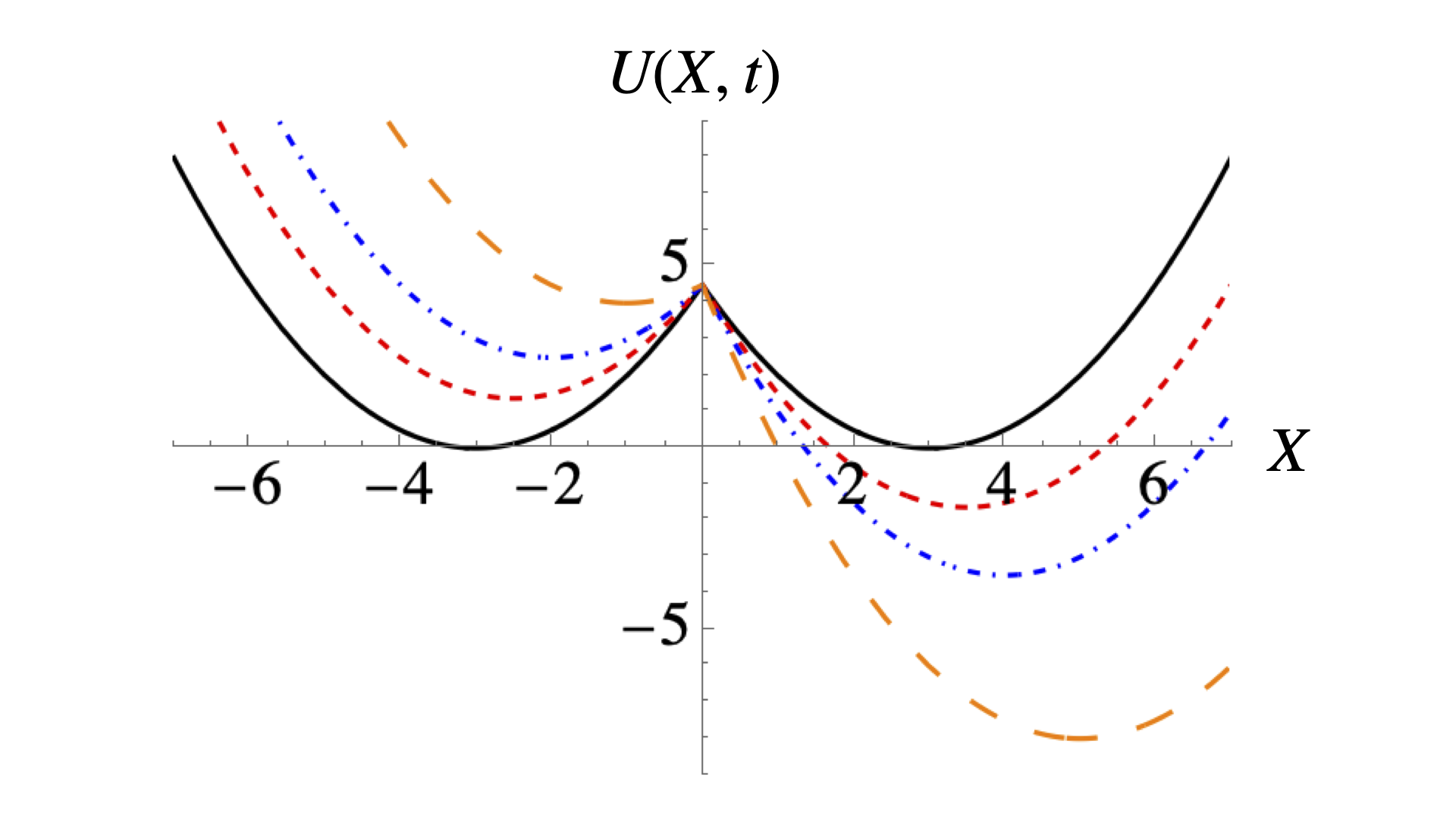}
\caption{\textcolor{black}{Double-parabolic} potential in dimensionless variables at various times with $a=3$.  The solid line is the potential at the beginning and end of the erasure protocol.  The short-dashed, dashed-dotted and long-dashed lines are the potential at intermediate times $t=50,\,100,\,200$, respectively.}
\label{Fig1}
\end{center}
\end{figure}

A concrete example, which we study in detail in Section~\ref{sec:F(t).lin},
\textcolor{black}{%
is a double-well potential constructed from two harmonic traps,
\begin{equation}
\label{eq:U}
U(X) = \frac{K}{2}[X - \textrm{sign}(X) a]^2
\, ,
\end{equation}
and a linear-in-time forcing protocol implemented as
\begin{subequations}
\label{eq:F(t).lin}
\begin{align}
\label{eq:F(t)erase.lin}
F_{\mathrm{erase}}(t) &= \lambda t \qquad \text{and}
\,  \\
F_{\mathrm{reset}}(t) &= \frac{\lambda \tau^*}{\tau-\tau^*}(\tau-t)
\, ,
\label{eq:F(t)reset.lin}
\end{align}
\end{subequations}
with $\lambda = Ka/\tau^* > 0$ (cf.~condition \eqref{eq:F(t=tau*)});
the constant $K$ is the curvature of the harmonic trap and $\textrm{sign}(\cdot)$ denotes the sign function.
}%
The corresponding change in the potential \eqref{eq:U} induced by $F_{\mathrm{erase}}(t)$
during the erasure phase is illustrated in Fig. \ref{Fig1}.

The evolution of the particle within the double-well potential is given by an overdamped non-autonomous Langevin equation
\cite{gardiner},
\begin{equation}
\label{eq:LE}
\gamma\dot{X}(t) = \textcolor{black}{-U'(X(t))}+ F(t) + \sqrt{2 \kB T \gamma} \, \xi(t)
\, ,
\end{equation}
where $\xi(t)$ is an unbiased Gaussian white noise source with correlations $\langle \xi(t)\xi(s)\rangle=\delta(t-s)$.
The \textcolor{black}{curvature of the potential $K=U''(\pm a)$ around the minima}, or the ``trap stiffness'', and the viscous friction coefficient $\gamma$ of the particle define the time scale $\gamma/K$, characterizing relaxation processes \emph{within} each potential well.

As noted above, we assume that $\gamma/K$ represents the fastest time scale in the deterministic part of the stochastic system,  and thus the erasure force $F(t)$ evolves on time scales much slower than $\gamma/K$.  Therefore, a particle ``equilibrates'' rapidly within a potential well, thereby guaranteeing that a new memory state is rapidly occupied when being updated by an externally imposed protocol.
\textcolor{black}{Despite this rapid ``local equilibration'' the memory is far from ``global equilibrium'' conditions, and close-to-equilibrium concepts, like the fluctuation-dissipation theorem \cite{Kubo1966}, are not obeyed in general.}
The intra-well relaxation time scale $\gamma/K$ can be directly controlled experimentally by adjusting the trap stiffness $K$ \cite{berut,berut1,dago,dago1}, so that our fast-relaxation assumption is hardly restrictive, and hence is a reasonable prerequisite for an effective realization of memory \cite{note0}.  We show below that the associated ``local equilibrium'' behavior within the individual potential wells is essential for calculating the energetic costs of the erasure process (see also the discussion at the beginning of \S~\ref{sec:dynamics}). We note here that the opposite case, of external forces $F(t)$ varying much faster than $\gamma/K$, corresponds to ``instantaneous'' manipulations of the memory potential and can be treated accordingly; we motivate the corresponding physical arguments in the context of an ``instantaneous'' resetting procedure in the discussion surrounding  \eqref{eq:quickReset}.  


We use the time ${\gamma}/{K}$, and the length $\sqrt{{\kB T}/{K}}$, to non-dimensionalize the Langevin equation \eqref{eq:LE}, which, in the same notation as the original variables, becomes
\textcolor{black}{%
\begin{equation}
\dot{X}(t) = -U'(X(t)) + F(t)+\sqrt{2} \, \xi(t)
\, ,
\label{lang}
\end{equation}
where the dimensionless force $F(t)$ is in units of $\sqrt{\kB T\,K}$ and the dimensionless potential $U(X)$ in units of $\kB T$.  The dimensionless condition that the potential barrier is much larger than the thermal energy is $U(0)-U(\pm a) \gg 1$,
and the dimensionless condition that the erasure force varies much more slowly than the relaxation of particles in the potential wells is $\dot{F}(t) \ll 1$.
From the dimensionless version of \eqref{eq:F(t=tau*)} we estimate a typical rate of change of the erasure protocol as $\dot{F}_{\mathrm{erase}}(t) \approx [U(0)-U(\pm a)]/a$. Therefore, our theory of the erasure process is valid when the combined constraint on the system parameters is such that $1 \ll U(0)-U(\pm a) \ll a\tau^*$.
}%
\subsection{Heat and work}
The central observables are the heat  dissipated into the thermal bath $Q$, and the total work received by the system during the erasure procedure $W$.  Stochastic energetics \cite{Sekimoto} and thermodynamics \cite{jarzynski,seifert,broeck1,seifert1} provide an established framework for evaluating such quantities along individual particle trajectories---taking into account the influence of thermal fluctuations.  Therefore, we can quantify the distributions resulting from many independent realizations of the erasure process.  In our case of a symmetric memory, the average system energy (for the overdamped dynamics of Eq. \eqref{eq:LE} this corresponds to the total potential energy \cite{Sekimoto}) is the same at the beginning and at the end of the erasure process, implying that the average work is exactly compensated by the average heat dissipated.  Therefore, since they sum to zero, it is sufficient to consider one of these two quantities to fully characterize the thermodynamics of the erasure process.  Here we focus on the total work exerted on the system.

The work is the change in the system energy driven by external forcing integrated along the particle trajectory $X(t)$ \cite{Sekimoto}.  Since the only external forcing of the system potential arises 
from the erasure protocol with potential energy $-F(t)X$,  we have
\begin{align}
\label{eq:W}
W
& = -\int_0^{\tau} \mathrm{d}t \, \dot{F}(t) X(t)
\nonumber \\
& = -\int_0^{\tau^*} \mathrm{d}t \, \dot{F}_{\mathrm{erase}}(t) X(t) - \int_{\tau^*}^{\tau} \mathrm{d}t \, \dot{F}_{\mathrm{reset}}(t) X(t)
\, .
\end{align}
Our main goal is to characterize the distribution of $W$ as a function of the system parameters.  Of particular relevance is the duration of the erasure process,  $\tau^*$.
\subsection{Typical system dynamics}
\label{sec:dynamics}
\begin{figure}
\begin{center}
\includegraphics[width=0.9\linewidth]{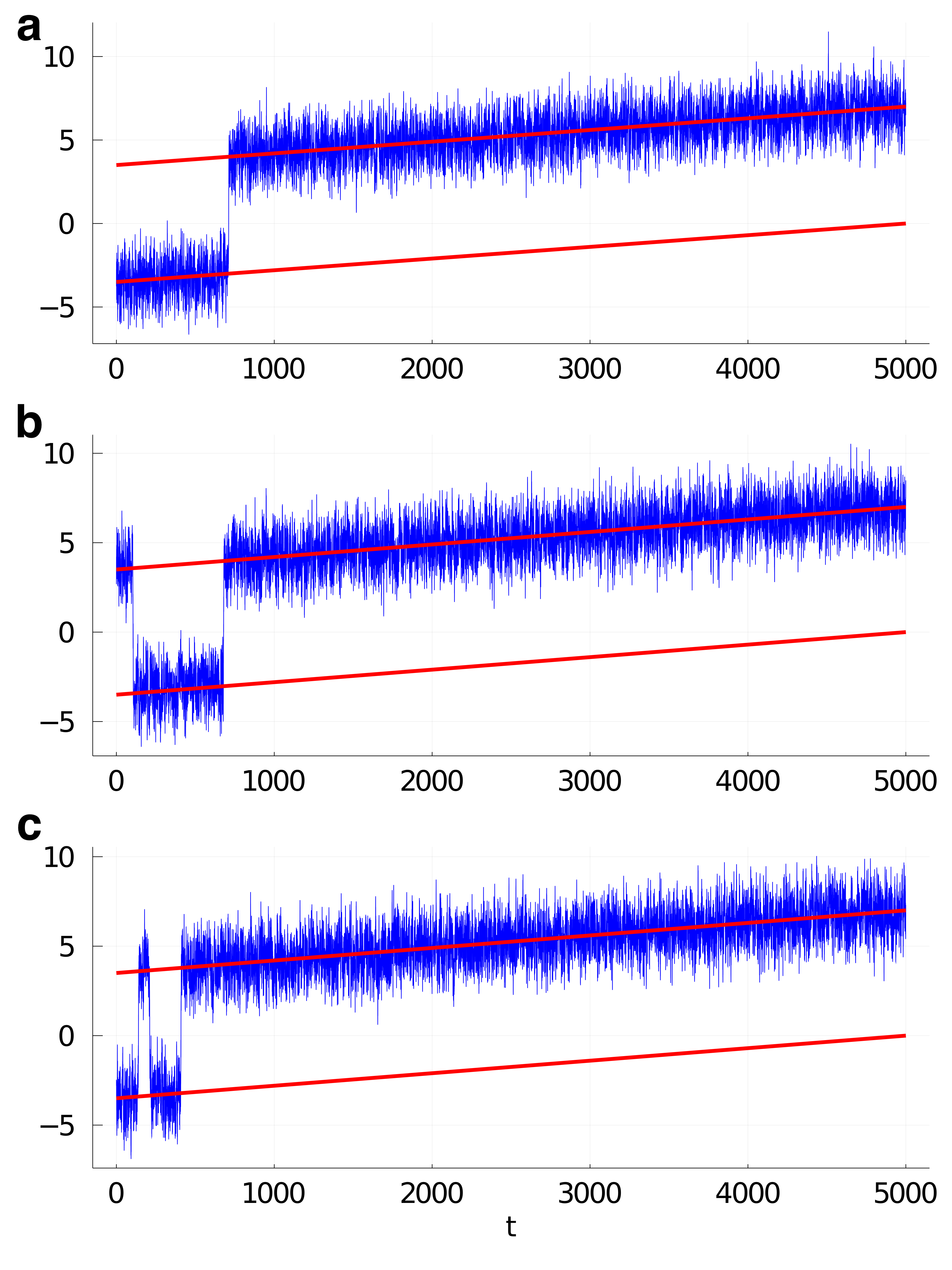}
\caption{Examples of trajectories (fluctuating blue lines) with one (a), two (b) and three (c)
transitions between the two potential wells, together with the two minima of the potential (straight red lines), as a function of time.
The upper (lower) straight red line in each panel corresponds to state 1 (state 0).
For all plots the parameter values are $a=3.5$ and $\tau^*=5000$.}
\label{Fig2}
\end{center}
\vspace{-1  cm}
\end{figure}

In Fig.~\ref{Fig2}, for the linear erasure protocol \eqref{eq:F(t).lin},  we show three typical trajectories of a Brownian particle from numerical simulations of the dimensionless Langevin equation \eqref{lang}.
We see the principal consequence of both the fast relaxation within the potential wells and the high barrier separating them; (i) despite the slow drift of the potential, a particle is mainly located near one of the potential minima in ``local equilibrium'' and (ii) occasionally, the particle transitions between the potential wells in an ``instantaneous'' jump.  Our erasure protocol in this example forces the particle to end up in state 1, but if it begins in state 0 (state 1), then in total it will experience an odd (even) number of jumps, as illustrated in Fig.~\ref{Fig2} for 1, 2 and 3 jumps.  Note that, for the parameters used here, observing more than 3 jumps is quite exceptional. Clearly,  these jumps are sufficiently rare, and the location of the potential wells shifts significantly between consecutive jumps, so that ``global equilibrium conditions'' are not realized.  Importantly, although our erasure protocol, $F(t)$, varies slowly relative to relaxation within the potential wells, it is still far from being ``quasi-statically slow'' and hence will not achieve the Landauer bound.

Given these results, we  adopt the following strategy to compute the energetic costs of information erasure in the double-well potential under non-quasi-static conditions.  As noted above, our key observable is the work performed on the system in order to erase one bit of information. We calculate this work under the assumptions that (a) the particle is in ``local equilibrium'' within a potential well and (b) the work performed during the ``instantaneous'' transitions from one potential well to the other is negligibly small (\S~\ref{sec:work.av}).  We quantify the contributions to the total work from the individual wells by analyzing the statistics of the inter-well jumps.
In \S~\ref{sec:F(t).lin} we explicitly demonstrate this general approach for the linear erasure protocol \eqref{eq:F(t).lin} \textcolor{black}{applied to the double-parabolic potential \eqref{eq:U}}, for which we calculate all relevant quantities analytically.

\section{The work distribution: General framework}
\label{sec:work}
We characterize the work distribution by determining its moments. The mean and the variance of the work are the most practical quantities.   Thus we now detail the general framework by calculating
$\langle W \rangle$ and $\Var(W)=\langle W^2 \rangle-\langle W \rangle^2$, where $\langle \cdot \rangle$ denotes the ensemble average over all possible trajectories.  It should then be clear how to extend our approach to higher moments.

\subsection{The Average Work}
\label{sec:work.av}
The average work follows directly from Eq. \eqref{eq:W} as
\begin{equation}
\langle W \rangle
= - \int_0^{\tau^*} \mathrm{d}t \, \dot{F}_{\mathrm{erase}}(t) \langle X(t) \rangle
	- \int_{\tau^*}^{\tau} \mathrm{d}t \, \dot{F}_{\mathrm{reset}}(t)\langle X(t)\rangle
\, .
\label{work}
\end{equation}
However,  because transitions between the potential wells can occur at any time, direct evaluation of $\langle X(t) \rangle$ is challenging, particularly during the first phase of the erasure protocol.  Indeed, whereas 
we know that the particle is in the right well (state 1) for $\tau^* \leq t \leq \tau$, we do not know within which potential well the particle is located at a particular time in $0 \leq t \leq \tau^*$.  In contrast, within a specific potential well the averages are straightforward to calculate.

Our approach is to collect all particle trajectories that ``instantaneously'' jump from one well to the other at specific times $t_i$ with $0 \leq t_1 < t_2 < \ldots< t_i \ldots < t_n \leq \tau^*$.  We remind the reader that the number of jumps $n$ is odd (even) when the trajectories begin in state 0 (state 1).  Thus we can decompose the ensemble average $\langle \cdot \rangle$ into an average over the sub-ensemble of trajectories with fixed transition times
$0 \leq t_1 < t_2 < \ldots< t_i \ldots < t_n \leq \tau^*$, and an average over the distribution of these transition times.
For the sub-ensemble, we calculate the average particle position within any of the time intervals $[t_i,t_{i+1}]$ by direct solution of the Langevin equation \eqref{lang}.  Applying the fast-relaxation assumption we then find that the average particle position
\textcolor{black}{is determined by the 
local ``quasi-equilibrium'' distribution for the current value of the erasure force within the potential well in which the particle resides during $[t_i,t_{i+1}]$, viz.
\begin{subequations}
\begin{align}
\label{eq:intra}
\langle X(t) \rangle_{\pm}
& \simeq \frac{\int_{-\infty}^{+\infty} \mathrm{d}X \, X \, e^{-\left[ V(a \mp X) - F_{\mathrm{erase}}(t)X \right]}}
			  {\int_{-\infty}^{+\infty} \mathrm{d}X \, e^{-\left[ V(a \mp X) - F_{\mathrm{erase}}(t)X \right]}}
\nonumber \\[2ex]
& = \pm \frac{\int_{-\infty}^{+\infty} \mathrm{d}X \, (a-X) e^{-\left[ V(X) \mp F_{\mathrm{erase}}(t)(a-X) \right]}}
			 {\int_{-\infty}^{+\infty} \mathrm{d}X \, e^{-\left[ V(X) \mp F_{\mathrm{erase}}(t)(a-X) \right]}}
\nonumber \\
& = \pm a + \frac{\partial}{\partial F_{\mathrm{erase}}(t)} \ln Z_\pm(F_{\mathrm{erase}}(t))
\, ,
\end{align}
with the partition functions
\begin{equation}
\label{eq:Z+-}
Z_\pm(F_{\mathrm{erase}}(t)) = \int_{-\infty}^{+\infty} \mathrm{d}x \, e^{-\left[ V(x) \pm F_{\mathrm{erase}}(t)x \right]}
\, .
\end{equation}
\end{subequations}
The plus (minus) sign in $\langle X(t) \rangle_{\pm}$ and $Z_\pm(F_{\mathrm{erase}}(t))$ refers to the right (left) well.  Note that the abbreviation $\langle . \rangle_{\pm}$
}%
denotes the average over all trajectories located \emph{within a specific potential well}
for the entire time interval over which the average is taken, and $\simeq$ denotes the leading-order contribution in the fast-relaxation approximation.

We can now determine the sub-ensemble average in the first integral of Eq.~\eqref{work}
by evaluating integrals over the time intervals $[t_i,t_{i+1}]$ between successive jumps as
\textcolor{black}{%
\begin{align}
\label{eq:phase1}
& -\int_0^{\tau^*} \mathrm{d}t \, \dot{F}_{\mathrm{erase}}(t) \langle X(t) \rangle
\nonumber \\
& \quad
= -\left\langle \sum_{i=0}^n \int_{t_i}^{t_{i+1}} \mathrm{d}t \, \dot{F}_{\mathrm{erase}}(t) X(t) \right\rangle
\nonumber \\
& \quad
= -\left\langle \sum_{i=0}^n \int_{t_i}^{t_{i+1}} \mathrm{d}t \, \dot{F}_{\mathrm{erase}}(t) \langle X(t) \rangle_{(-)^{i+n}} \right\rangle_{\!\!\{t_i\}}
\nonumber \\
& \quad
\simeq -\left\langle \sum_{i=0}^n \int_{F_{\mathrm{erase}}(t_i)}^{F_{\mathrm{erase}}(t_{i+1})} \mathrm{d}F_{\mathrm{erase}}(t)
\right.
\nonumber \\
& \quad\quad \mbox{} \times \left.
	\left[ (-1)^{i+n} a + \frac{\partial}{\partial F_{\mathrm{erase}}(t)} \ln Z_{(-)^{i+n}}(F_{\mathrm{erase}}(t)) \right]
\right\rangle_{\!\!\{t_i\}}
\nonumber \\[1ex]
& \quad
= a \left\langle \sum_{i=0}^n (-1)^{i+n}\left[F_{\mathrm{erase}}(t_i) - F_{\mathrm{erase}}(t_{i+1})\right]
\right.
\nonumber \\
& \qquad\qquad \mbox{}
	+ \left. \ln \frac{Z_{(-)^{i+n}}(F_{\mathrm{erase}}(t_i))}{Z_{(-)^{i+n}}(F_{\mathrm{erase}}(t_{i+1}))}
 \right\rangle_{\!\!\{t_i\}}
\nonumber \\
& \quad
= 2a \left\langle \sum_{i=1}^n (-1)^{i+n} F_{\mathrm{erase}}(t_i) \right\rangle_{\!\!\{t_i\}}
	- F_{\mathrm{erase}}(\tau^*) a
\nonumber \\[1ex]
& \qquad\qquad
+ \left\langle 
	\sum_{i=0}^n \ln \frac{Z_{(-)^{i+n}}(F_{\mathrm{erase}}(t_i))}{Z_{(-)^{i+n}}(F_{\mathrm{erase}}(t_{i+1}))}
\right\rangle_{\!\!\{t_i\}}
\, ,
\end{align}
}%
where we let $t_0=0$ and $t_{n+1}=\tau^*$ to simplify notation. 
Here, for each time interval $[t_i,t_{i+1}]$,  $t_i$ refers to the time ``just after'' the jump into the current potential well, and $t_{i+1}$ to the time ``just before'' the next jump.  The average over the distributions of transition
times for all possible $n$ is denoted by the subscript $\{t_i\}$.

The second integral in Eq. \eqref{work} is the average work from the resetting phase of the erasure protocol.   As described above, at the end of the ``erasure phase'' the particle is in the right well (state 1).  Now, to prevent the particle from jumping back to state 0 during the ``resetting'' phase, it is desirable to minimize the resetting interval, which at the very least should be much shorter than the first phase of the protocol so that $\tau-\tau^* \ll \tau^*$.
However, employing the same reasoning as above, we must ensure that the reset dynamics is still slow relative to the relaxation processes within the potential wells, $\dot{F}_{\mathrm{reset}}(t) \ll 1$.  This imposes a stronger constraint, $(\tau-\tau^*)/\tau^* \ll 1$, on the rate of change of the external forcing protocol.  Under these conditions, the average particle position follows the potential minimum of state 1, and we can evaluate the second term in Eq. \eqref{work} similarly to that of the first term, which gives
\textcolor{black}{%
\begin{align}
\label{eq:phase2}
& - \int_{\tau^*}^{\tau} \mathrm{d}t \, \dot{F}_{\mathrm{reset}}(t)\langle X(t)\rangle
= - \int_{\tau^*}^{\tau} \mathrm{d}t \, \dot{F}_{\mathrm{reset}}(t)\langle X(t)\rangle_+
\nonumber \\
& \quad
= -\int_{\tau^*}^{\tau} \mathrm{d}t \, \dot{F}_{\mathrm{reset}}(t)
	\left[ a + \frac{\partial}{\partial F_{\mathrm{erase}}(t)} \ln Z_+(F_{\mathrm{erase}}(t)) \right]
\nonumber \\[1ex]
& \quad
= F_{\mathrm{reset}}(\tau^*) a
	+ \ln \frac{Z_+(F_{\mathrm{erase}}(\tau^*))}{Z_+(F_{\mathrm{erase}}(\tau))}
\nonumber \\[1ex]
& \quad
= F_{\mathrm{reset}}(\tau^*) a
	+ \ln \frac{Z_+(F_{\mathrm{erase}}(\tau^*))}{Z_+(F_{\mathrm{erase}}(0))}
\, ,
\end{align}
where, from condition \eqref{eq:F(t=0)}, we used $F_{\mathrm{reset}}(\tau)=F_{\mathrm{reset}}(0)$. Finally, we combine Eqs. \eqref{eq:phase1} and \eqref{eq:phase2} to give the total average work as
\begin{align}
\label{eq:<W>}
& \langle W \rangle = 2a \left\langle \sum_{i=1}^n (-1)^{i+n} F_{\mathrm{erase}}(t_i) \right\rangle_{\!\!\{t_i\}}
\nonumber \\
& \mbox{}
	+ \left\langle
		\sum_{i=0}^n \ln \left[ \frac{Z_{(-)^{i+n}}(F_{\mathrm{erase}}(t_i))}{Z_{+}(F_{\mathrm{erase}}(t_i))}
								\frac{Z_{+}(F_{\mathrm{erase}}(t_{i+1}))}{Z_{(-)^{i+n}}(F_{\mathrm{erase}}(t_{i+1}))}
						 \right]
	\right\rangle_{\!\!\{t_i\}}
\, .
\end{align}
We note that the average work depends on the shape of the tilted double-well potential through the statistics of the jump times $\{t_i\}$ (see also \S~\ref{sec:survival}) and through the partition functions $Z_\pm(F_{\mathrm{erase}}(t_{i}))$, Eq.~\eqref{eq:Z+-}.
If $V(X)$ is symmetric, i.e. $V(-X)=V(X)$, the latter dependence drops out, because $Z_-(F_{\mathrm{erase}}(t))=Z_+(F_{\mathrm{erase}}(t))$.
For mirror-symmetric potential wells like the one in Fig.~\ref{Fig1}, we thus find
\begin{equation}
\label{eq:<W>sym}
\langle W \rangle = 2a \left\langle \sum_{i=1}^n (-1)^{i+n} F_{\mathrm{erase}}(t_i) \right\rangle_{\!\!\{t_i\}}
\, .
\end{equation}
}%

There is a finite, albeit very small, probability that the particle might jump to state 0 during the resetting phase, $F_{\mathrm{reset}}(t)$.  To avoid such imperfect erasure processes, we control the external force by switching it from $F(\tau^*)=F_{\mathrm{erase}}(\tau^*)$ to $F(\tau)=0$ over an infinitesimal time interval, such that practically $\tau = \tau^*$.  Thus, the particle does not move while the potential is reset to its initial configuration, and the work performed corresponds to the change in potential energy at a fixed particle position, allowing us to write the average as
\begin{align}
\label{eq:quickReset}
& - \int_{\tau^*}^{\tau} \mathrm{d}t \, \dot{F}_{\mathrm{reset}}(t)\langle X(t)\rangle
\nonumber \\[0ex]
& \qquad\qquad
= \left\langle U(X(\tau)) - F_{\mathrm{reset}}(\tau)X(\tau) \right\rangle
\nonumber \\[0.5ex]
& \qquad\qquad\qquad \mbox{}
	- \left\langle U(X(\tau^*)) - F_{\mathrm{reset}}(\tau^*)X(\tau^*) \right\rangle
\nonumber \\[1ex]
& \qquad\qquad
= F_{\mathrm{reset}}(\tau^*) \langle X(\tau^*) \rangle_+
\, .
\end{align}
\textcolor{black}{%
The subsequent relaxation of the particle towards the minimum does not contribute to the work,
but is instead dissipated as heat into the thermal bath.  
To compare the instantaneous work \eqref{eq:quickReset}
with Eq. \eqref{eq:phase2} for the average work in the resetting phase of the erasure protocol, we
evaluate their difference using partial integration,
$F_{\mathrm{reset}}(\tau^*) \langle X(\tau^*) \rangle_+ + \int_{\tau^*}^{\tau} \mathrm{d}t \, \dot{F}_{\mathrm{reset}}(t)\langle X(t)\rangle_+
= -\int_{\tau^*}^{\tau} \mathrm{d}t \, F_{\mathrm{reset}}(t)\tfrac{\mathrm{d}}{\mathrm{d}t}\langle X(t)\rangle_+ > 0$.
Positivity follows because $F_{\mathrm{reset}}(t)>0$ for all $t>0$, which implies that the average particle position decreases during the resetting phase.
}%
Hence, when instantaneously resetting the erasure protocol the total work required to erase one bit of information is larger than that in \eqref{eq:<W>} by an amount \textcolor{black}{$-\int_{\tau^*}^{\tau} \mathrm{d}t \, F_{\mathrm{reset}}(t)\tfrac{\mathrm{d}}{\mathrm{d}t}\langle X(t)\rangle_+$}. This is because switching off the erasure force during a finite time interval $(\tau^*,\tau]$ using
$F_{\mathrm{reset}}(t)$ where $\dot{F}_{\mathrm{reset}}(t) \ll 1$, while the particle is solely confined to state 1, corresponds to a quasi-static process under
local equilibrium conditions within the potential well representing state 1.  In contrast, the instantaneous resetting is ``maximally far'' from local equilibrium conditions.

\subsection{The Variance of the Work}

Using the same approach we used to calculate the average work in Eq. \eqref{eq:phase1}, we can evaluate higher moments. We focus on the variance and begin by determining the second moment
of the work exerted during the first phase of the erasure protocol, which is
\begin{widetext}
\begin{align}
& \left\langle \left( \int_0^{\tau^*} \mathrm{d}t \, \dot{F}_{\mathrm{erase}}(t) X(t) \right)^{\!\!2} \right\rangle
= \left\langle \left( \sum_{i=0}^n \int_{t_i}^{t_{i+1}} \mathrm{d}t \, \dot{F}_{\mathrm{erase}}(t) X(t) \right)^{\!\!2} \right\rangle
\nonumber \\
& \qquad
= \left\langle \sum_{i=0}^n \int_{t_i}^{t_{i+1}} \mathrm{d}t \,\dot{F}_{\mathrm{erase}}(t) \int_{t_i}^{t_{i+1}} \mathrm{d}s \, \dot{F}_{\mathrm{erase}}(s)
	\Big( \langle X(t)X(s) \rangle_{(-)^{i+n}} - \langle X(t) \rangle_{(-)^{i+n}} \langle X(s) \rangle_{(-)^{i+n}} \Big)
\right\rangle_{\!\!\{t_i\}}
\nonumber \\
& \qquad\qquad\qquad \mbox{}
+ \left\langle 
	\left( \sum_{i=0}^n \int_{t_i}^{t_{i+1}} \mathrm{d}t \, \dot{F}_{\mathrm{erase}}(t) \langle X(t) \rangle_{(-)^{i+n}} \right)^{\!\!2}
\right\rangle_{\!\!\{t_i\}}
\, .
\end{align}
\end{widetext}
As in Eq. \eqref{eq:phase1}, we split the ensemble average into an average over the fluctuations within a specific potential well, $\langle \cdot \rangle_{\pm}$, and an average over the distribution of jump times from one well to the other, $\langle \cdot \rangle_{\{t_i\}}$.  Additionally, we assume that the fluctuations within different potential wells are uncorrelated.
\textcolor{black}{
In the first term, we solve the Langevin equation \eqref{lang} to calculate the autocorrelation $\langle X(t)X(s) \rangle_{\pm} - \langle X(t) \rangle_{\pm} \langle X(s) \rangle_{\pm}$ within a given potential well.  Under our fast-relaxation assumption, to lowest order we obtain 
$\langle X(t)X(s) \rangle_{\pm} - \langle X(t) \rangle_{\pm} \langle X(s) \rangle_{\pm} \simeq \delta(t-s)\,\tfrac{\partial^2}{\partial F_{\mathrm{erase}}(t)^2} Z_\pm(F_{\mathrm{erase}}(t))$, and hence the first term can be evaluated as $\left\langle \sum_{i=0}^n \int_{t_i}^{t_{i+1}} \mathrm{d}t \,\dot{F}_{\mathrm{erase}}^2(t)\tfrac{\partial^2}{\partial F_{\mathrm{erase}}(t)^2} Z_{(-)^{i+n}}(F_{\mathrm{erase}}(t)) \right\rangle_{\!\!\{t_i\}}$.
In the second term, we use the result
$-\sum_{i=0}^n \int_{t_i}^{t_{i+1}} \mathrm{d}t \, \dot{F}_{\mathrm{erase}}(t) \langle X(t) \rangle_{(-)^{i+n}} =
2a \sum_{i=1}^n (-1)^{i+n} F_{\mathrm{erase}}(t_i) - a F_{\mathrm{erase}}(\tau^*) + \sum_{i=0}^n \ln \tfrac{Z_{(-)^{i+n}}(F_{\mathrm{erase}}(t_i))}{Z_{(-)^{i+n}}(F_{\mathrm{erase}}(t_{i+1}))}$,
which was derived above (see the development in Eq. \eqref{eq:phase1}).  Therefore, we find that the variance of the work during the first phase of the erasure protocol is
\begin{widetext}
\begin{align}
\label{eq:varW1}
& \left\langle \left( \int_0^{\tau^*} \mathrm{d}t \, \dot{F}_{\mathrm{erase}}(t) X(t) \right)^{\!\!2} \right\rangle
- \left\langle \int_0^{\tau^*} \mathrm{d}t \, \dot{F}_{\mathrm{erase}}(t) X(t) \right\rangle^{\!\!2}
\simeq \left\langle \sum_{i=0}^n \int_{t_i}^{t_{i+1}} \mathrm{d}t \,\dot{F}_{\mathrm{erase}}^2(t)
			\frac{\partial^2}{\partial F_{\mathrm{erase}}(t)^2} Z_{(-)^{i+n}}(F_{\mathrm{erase}}(t)) \right\rangle_{\!\!\{t_i\}}
\nonumber \\ & \qquad\qquad\qquad\qquad\qquad\qquad\qquad\qquad\qquad \mbox{}
+ \left\langle \left( 2a \sum_{i=1}^n (-1)^{i+n} F_{\mathrm{erase}}(t_i)
	+ \sum_{i=0}^n \ln \frac{Z_{(-)^{i+n}}(F_{\mathrm{erase}}(t_i))}{Z_{(-)^{i+n}}(F_{\mathrm{erase}}(t_{i+1}))} \right)^{\!\!2}
  \right\rangle_{\!\!\{t_i\}}
\nonumber \\ & \qquad\qquad\qquad\qquad\qquad\qquad\qquad\qquad\qquad \mbox{}
- \left\langle  2a \sum_{i=1}^n (-1)^{i+n} F_{\mathrm{erase}}(t_i)
	+ \sum_{i=0}^n \ln \frac{Z_{(-)^{i+n}}(F_{\mathrm{erase}}(t_i))}{Z_{(-)^{i+n}}(F_{\mathrm{erase}}(t_{i+1}))}
  \right\rangle_{\!\!\{t_i\}}^{\!\!2}.
\end{align}
\end{widetext}
}

We use a similar approach to determine the variance of the work during the resetting phase.  This is simpler, because the particle stays in state 1 for all $t \in (\tau^*,\tau]$, and we find
\textcolor{black}{
\begin{multline}
\label{eq:varW2}
\left\langle \left( \int_{\tau^*}^{\tau} \mathrm{d}t \, \dot{F}_{\mathrm{reset}}(t) X(t) \right)^{\!\!2} \right\rangle
- \left\langle \int_{\tau^*}^{\tau} \mathrm{d}t \, \dot{F}_{\mathrm{reset}}(t) X(t) \right\rangle^{\!\!2}
\\
= \int_{\tau^*}^{\tau} \mathrm{d}t \,\dot{F}_{\mathrm{reset}}^2(t) \frac{\partial^2}{\partial F_{\mathrm{erase}}(t)^2} \ln Z_{+}(F_{\mathrm{erase}}(t))
\, .
\end{multline}
}

We combine the two parts of the erasure protocol, Eqs. \eqref{eq:varW1} and \eqref{eq:varW2}, with $t \in [0,\tau^*]$ and $t \in (\tau^*,\tau]$, under the
assumption that these two phases are uncorrelated, giving the variance of the total work, $\Var(W) = \langle W^2 \rangle - \langle W \rangle^2$,  as
\textcolor{black}{
\begin{widetext}
\begin{align}
\label{eq:var(W)}
\Var(W) & =
\left\langle \sum_{i=0}^n \int_{t_i}^{t_{i+1}} \mathrm{d}t \,\dot{F}_{\mathrm{erase}}^2(t)
	\frac{\partial^2}{\partial F_{\mathrm{erase}}(t)^2} \ln Z_{(-)^{i+n}}(F_{\mathrm{erase}}(t))
\right\rangle_{\!\!\{t_i\}}
+ \int_{\tau^*}^{\tau} \mathrm{d}t \,\dot{F}_{\mathrm{reset}}^2(t) \frac{\partial^2}{\partial F_{\mathrm{erase}}(t)^2} \ln Z_{+}(F_{\mathrm{erase}}(t))
\nonumber \\ & \qquad\qquad \mbox{}
+ \Var \left( 2a \sum_{i=1}^n (-1)^{i+n} F_{\mathrm{erase}}(t_i)
	+ \sum_{i=0}^n \ln \frac{Z_{(-)^{i+n}}(F_{\mathrm{erase}}(t_i))}{Z_{(-)^{i+n}}(F_{\mathrm{erase}}(t_{i+1}))} \right)
\nonumber \\
& =
\left\langle \sum_{i=0}^n \int_{t_i}^{t_{i+1}} \mathrm{d}t \,\dot{F}_{\mathrm{erase}}^2(t)
	\frac{\partial^2}{\partial F_{\mathrm{erase}}(t)^2} \ln Z_{(-)^{i+n}}(F_{\mathrm{erase}}(t))
\right\rangle_{\!\!\{t_i\}}
+ \int_{\tau^*}^{\tau} \mathrm{d}t \,\dot{F}_{\mathrm{reset}}^2(t) \frac{\partial^2}{\partial F_{\mathrm{erase}}(t)^2} \ln Z_{+}(F_{\mathrm{erase}}(t))
\nonumber \\ & \qquad\qquad \mbox{}
+ \Var \left( 2a \sum_{i=1}^n (-1)^{i+n} F_{\mathrm{erase}}(t_i)
	+ \sum_{i=0}^n \ln \left[ \frac{Z_{(-)^{i+n}}(F_{\mathrm{erase}}(t_i))}{Z_{+}(F_{\mathrm{erase}}(t_i))}
							  \frac{Z_{+}(F_{\mathrm{erase}}(t_{i+1}))}{Z_{(-)^{i+n}}(F_{\mathrm{erase}}(t_{i+1}))}
					   \right]
\right)
\, .
\end{align}
\end{widetext}
After adding the $\{t_i\}$-independent constant
$\ln \frac{Z_+(F_{\mathrm{erase}}(\tau^*))}{Z_+(F_{\mathrm{erase}}(0))}
= \sum_{i=0}^n \ln \frac{Z_{+}(F_{\mathrm{erase}}(t_{i+1}))}{Z_{+}(F_{\mathrm{erase}}(t_i))}$,
we see that the second term
is the variance of the result we found for the average work, Eq. \eqref{eq:<W>}; it is related to the variance of the jump-time distribution.
Moreover, there is an additional term involving $\dot{F}^2(t)$, which we expect to be negligibly small relative to the second term because $\dot{F}(t) \ll 1$. In the example of a linear erasure protocol applied to a double-parabolic potential (\S~\ref{sec:F(t).lin:W}) we show that this first term is related to the variance accumulated from the fluctuating motion within the potential wells.
Indeed, because
$Z_-(F_{\mathrm{erase}}(t))=Z_+(F_{\mathrm{erase}}(t))$ for mirror-symmetric potential wells $V(-X)=V(X)$, Eq.~\eqref{eq:var(W)} simplifies considerably to 
\begin{multline}
\label{eq:var(W)sym}
\Var(W)
= \int_{0}^{\tau} \mathrm{d}t \,\dot{F}^2(t) \frac{\partial^2}{\partial F_{\mathrm{erase}}(t)^2} \ln Z_{+}(F_{\mathrm{erase}}(t))
\\
+ 4a^4 \, \Var \!\left( \sum_{i=1}^n (-1)^{i+n} F_{\mathrm{erase}}(t_i) \right)
\, .
\end{multline}
}

The remaining averages $\langle \cdot \rangle_{\{t_i\}}$ in the mean work \eqref{eq:<W>} and its variance \eqref{eq:var(W)}
involve the statistics of the jump times $\{t_i\}$, and thus depend on the explicit form of the erasure protocol $F_{\mathrm{erase}}(t)$.
Now we describe a general strategy  to calculate these averages.

\subsection{Jump-time statistics}
\label{sec:P(t_i)}
The main task is to find the probability density function $\mathcal{P}_n(t_1,t_2,\ldots,t_n)$, or PDF,
for the distribution of jump times $\{t_i\}$ for any number of jumps $n$ that a particle may perform between the potential wells during the erasure phase of the protocol.  We express $\mathcal{P}_n(t_1,t_2,\ldots,t_n)$ in terms of the probabilities $P_{0 \rightleftarrows 1}(\underline{t},\overline{t})$ that the particle jumps from state 0 to state 1 at $t=\overline{t}$, given that it arrived in state 0 at $t=\underline{t}<\overline{t}$ (upper arrow in the subscript), and analogously that the particle jumps from state 1 to state 0 after having arrived
in state 1 at $t=\underline{t}<\overline{t}$ (lower arrow).
These are expressed as
\textcolor{black}{
\begin{equation}
\label{eq:PS}
P_{0 \rightleftarrows 1}(\underline{t},\overline{t}) = -\frac{\partial}{\partial \overline{t}}S_{0,1}(\underline{t},\overline{t})
\, ,
\end{equation}
}%
where $S_{0,1}(\underline{t},\overline{t})$ is the survival probability that the particle remains in state 0,1 from time $t=\underline{t}$ until time $t=\overline{t}>\underline{t}$.  The first (second) subscript in $S_{0,1}(\underline{t},\overline{t})$ is connected to the upper (lower) arrow in $P_{0 \rightleftarrows 1}(\underline{t},\overline{t})$. 
Given the ostensibly instantaneous character of the jumps, we will associate
\textcolor{black}{%
$S_{0,1}(\underline{t},\overline{t})$ with a transition rate between the two wells (see \S~\ref{sec:survival}).
}

\textcolor{black}{
The probability of $n$ jumps occurring
at specific times $0 \leq t_1 < t_2 < \ldots < t_n \leq \tau^*$ is
\begin{widetext}
\begin{equation}
\label{eq:P(t_i)}
\mathcal{P}_n(t_1,t_2,\ldots,t_n)
= \frac{1}{2} P_{0 \rightleftarrows 1}(0,t_1) P_{0 \leftrightarrows 1}(t_1,t_2) \dots P_{0\rightarrow1}(t_{n-1},t_n)S_1(t_n,\tau^*)
\, .
\end{equation}
\end{widetext}
The upper (lower) arrows in the subscripts refer to $n$ odd (even), with the particle starting in state 0 (state 1), so that $\mathcal{P}_0 = \tfrac{1}{2}S_1(0,\tau^*)$, and the prefactor $1/2$ accounts for the particle starting in either state with identical 50\% probability.
The normalization of $\mathcal{P}_n(t_1,t_2,\ldots,t_n)$ involves the probability $\Pi_n$ of observing exactly $n$ jumps within the time interval $[0,\tau^*]$ that takes into account all possible sequences of transition times $0 \leq t_1 < t_2 < \ldots < t_n \leq \tau^*$ as follows, 
\begin{widetext}
\begin{equation}
\label{eq:PIn}
\Pi_n = \frac{1}{2} \int_0^{\tau^*}\mathrm{d}t_1 P_{0 \rightleftarrows 1}(0,t_1)
	\int_{t_1}^{\tau^*}\mathrm{d}t_2 P_{0 \leftrightarrows 1}(t_1,t_2) \, \dots
	\int_{t_{n-1}}^{\tau^*}\mathrm{d}t_n P_{0\rightarrow1}(t_{n-1},t_n)S_1(t_n,\tau^*)
\, .
\end{equation}
\end{widetext}
The $\Pi_n$ are normalized with respect to an infinite number of inter-well jump transitions, $\sum_{n=0}^\infty \Pi_n = 1$.
}

However, recalling our discussion surrounding Fig.~\ref{Fig2}, we note that typically there are no more than a handful of transitions during the erasure phase.  Thus, $\Pi_n$ differs from zero only for small $n$, and we show in \S~\ref{sec:nfinite} that Eq. \eqref{eq:PIn} can be replaced by enforcing normalization for the relevant small number of inter-well transitions.
\textcolor{black}{%
In that case the probabilities $\Pi_n$ are given, and we write $\mathcal{P}_n(t_1,t_2,\ldots,t_n)$, Eq.~\eqref{eq:P(t_i)}, for any $n>0$ as
}%
\begin{widetext}
\begin{equation}
\label{eq:P(t_i)givenPIn}
\mathcal{P}_n(t_1,t_2,\ldots,t_n)
= \Pi_n
\frac{P_{0 \rightleftarrows 1}(0,t_1) P_{0 \leftrightarrows 1}(t_1,t_2) \dots P_{0\rightarrow1}(t_{n-1},t_n)S_1(t_n,\tau^*)}	
	{\int_0^{\tau^*}\mathrm{d}t_1 P_{0 \rightleftarrows 1}(0,t_1)
	\int_{t_1}^{\tau^*}\mathrm{d}t_2 P_{0 \leftrightarrows 1}(t_1,t_2) \, \dots
	\int_{t_{n-1}}^{\tau^*}\mathrm{d}t_n P_{0\rightarrow1}(t_{n-1},t_n)S_1(t_n,\tau^*)}
\, ,
\end{equation}
\end{widetext}
where, as in Eqs.~\eqref{eq:P(t_i)} and \eqref{eq:PIn}, the upper (lower) arrows in the subscripts refer to $n$ odd (even).
\textcolor{black}{%
The ratio guarantees normalization of $\mathcal{P}_n(t_1,t_2,\ldots,t_n)$ independently of the normalization of the $\Pi_n$.
}

Now we can use Eq.~\eqref{eq:P(t_i)}, or equivalently Eq.~\eqref{eq:P(t_i)givenPIn}, to calculate averages $\langle \cdot \rangle_{\{t_i\}}$ over the jump statistics as they appear in Eqs. \eqref{eq:<W>} and \eqref{eq:var(W)}; 
\begin{widetext}
\begin{equation}
\langle G(\{t_i\}) \rangle_{\{t_i\}}
= \sum_{n=1}^\infty \int_0^{\tau^*} \mathrm{d}t_1 \int_{t_1}^{\tau^*} \mathrm{d}t_2 \, \ldots \int_{t_{n-1}}^{\tau^*} \mathrm{d}t_n \,
		G(\{t_i\}) \mathcal{P}_n(t_1,t_2,\ldots,t_n)
\, ,
\end{equation}
\end{widetext}
where $G(\{t_i\})$ is an arbitrary function of the transition times $\{t_i\}$ and the number $n$ of jumps between potential wells within the interval $[0,\tau^*]$, but we must respect the time ordering $0 \leq t_1 < t_2 < \ldots < t_n \leq \tau^*$.

\subsection{Approximation for a finite number of jumps}
\label{sec:nfinite}

As the duration of the erasure protocol increases, so too does the typical number of transitions between the potential wells.  Moreover, as the potential tilt rate $F_{\mathrm{erase}}(t)$ slows, so too does the duration that exit rates from both potential minima remain comparable, and hence there are a larger number of jumps between states before the potential is sufficiently tilted that transitions from state 1 back to state 0 are strongly suppressed.

To determine the probabilities $\Pi_n$ for $n$ transitions occurring in the interval $[0,\tau^*]$, we focus on sufficiently rapid erasure protocols that no more than three jumps are observed, and the fast-relaxation requirement $\dot{F}_{\mathrm{erase}}(t) \ll 1$, as discussed in \S~\ref{sec:model}, is obeyed.  We treat the cases of $N=1$, $N=2$ or $N=3$ transitions individually, and our central approximation is to set the probability of observing more than $N$ transitions to zero, that is $\Pi_n = 0$ for $n>N$.  At the beginning of the protocol a particle can be found in either state with an equal probability of $1/2$, and at the end of the protocol in state 1
with probability $1$.  All $\Pi_n$ with $n$ even and all $\Pi_n$ with $n$ odd separately sum to $1/2$, whatever the total number $N$ of jumps we consider;
\begin{equation}
\sum_{n=0}^N \Pi_{2n} = \frac{1}{2}
\, , \qquad
\sum_{n=0}^N \Pi_{2n+1} = \frac{1}{2}
\, .
\end{equation}
Finally, for an even (odd) number of transitions to occur the particle must be in state 1 (state 0) at the beginning of the erasure process.

\underline{Total number of jumps: $N=1$}.  We approximate the probability $S_1(0,\tau^*)$ that a particle stays in state 1 when it starts in state 1 to be unity, and hence
we have
\begin{equation}
\label{eq:N1}
\begin{array}{rcl}
\Pi_0 & = & \frac{1}{2} \\[1ex]
\Pi_1 & = & \frac{1}{2} 
\end{array}
\qquad \mbox{($N=1$ transition).}
\end{equation}

\underline{Total number of jumps: $N=2$}. Here when a particle starts in state 1, it can either remain in that state for the entire duration of the first part of the erasure protocol, $[0,\tau^*]$, or it can jump once to state 0 and then back to state 1.  The probability for the first case is $S_1(0,\tau^*)$, and that for the second case is $1-S_1(0,\tau^*)$, and hence 
\begin{equation}
\label{eq:N2}
\begin{array}{rcl}
\Pi_0 & = & \frac{1}{2}S_1(0,\tau^*) \\[1ex]
\Pi_1 & = & \frac{1}{2}				 \\[1ex]
\Pi_2 & = & \frac{1}{2}[1-S_1(0,\tau^*)]
\end{array}
\qquad \mbox{($N=2$ transitions).}
\end{equation}

\underline{Total number of jumps: $N=3$}.  Here, the probabilities for an even number of jumps remain unchanged, because the only additional process is that with 3 jumps starting in state 0. This implies that the probability for just 1 transition is proportional to $\int_0^{\tau^*} \mathrm{d}t \, P_{0 \rightarrow 1}(0,t)S_1(t,\tau^*)$. Since our erasure protocol forces a particle starting in state 0 to jump to state 1 at some time within the interval $[0,\tau^*]$, the unknown proportionality constant is given by $\int_0^{\tau^*} \mathrm{d}t \, P_{0 \rightarrow 1}(0,t)$.  Therefore, we find
\begin{equation}
\label{eq:N3}
\begin{array}{rcl}
\Pi_0 & = & \frac{1}{2}S_1(0,\tau^*) 	 \\[0.5ex]
\Pi_1 & = & \frac{1}{2}\frac{\int_0^{\tau^*} \mathrm{d}t \, P_{0 \rightarrow 1}(0,t)S_1(t,\tau^*)}
							{\int_0^{\tau^*} \mathrm{d}t \, P_{0 \rightarrow 1}(0,t)}	\\[1.5ex]
\Pi_2 & = & \frac{1}{2}[1-S_1(0,\tau^*)] \\[1ex]
\Pi_3 & = & \frac{1}{2}- \Pi_1			 
\end{array}
\qquad \mbox{($N=3$ transitions).}
\end{equation}
Finally, we remark that the considerations leading to Eqs. \eqref{eq:N1}, \eqref{eq:N2}, and \eqref{eq:N3} can easily be extended to $N \geq 4$.

\subsection{Calculation of $S_{0,1}(\underline{t},\overline{t})$ and $P_{0 \rightleftarrows 1}(\underline{t},\overline{t})$}
\label{sec:survival}

The survival probabilities of a particle staying in state 0 (first subscript), or state 1 (second subscript), from time $\underline{t}$ to time $\overline{t}>\underline{t}$ are denoted $S_{0,1}(\underline{t},\overline{t})$.
\textcolor{black}{%
From $S_{0,1}(\underline{t},\overline{t})$ the probability $P_{0 \rightleftarrows 1}(\underline{t},\overline{t})$ can be derived with respect to the final time $\overline{t}$ (see also Eq.~\eqref{eq:PS}).
}

\textcolor{black}{%
The transitions between the two potential wells occur ``instantaneously'' relative to the rate of change of the erasure protocol, $\dot{F}(t)\ll 1$ for $t>0$.
So long as the potential barrier between the wells is much larger than the thermal energy, we can 
treat the inter-well jumps as rate-driven escape processes within a stationary potential,
which is formed by the current value of the erasure force $F_{\mathrm{erase}}(t)$. 
Hence, we write the survival probability in terms of the escape rates
$r_{0,1}(t)$ from state 0 (first subscript) or state 1 (second subscript),
\begin{equation}
\label{eq:S01}
S_{0,1}(\underline{t},\overline{t}) = \exp\left[ 
-\int_{\underline{t}}^{\overline{t}} \mathrm{d}t \, r_{0,1}(t) \right]
\, ,
\end{equation}
and calculate these rates using the Kramers formula \cite{kramers1940,hanggi1990};
\begin{equation}
\label{eq:rate_smoothMax}
r_{0,1}(t)=\frac{1}{2\pi}\sqrt{|U''(X_{\mathrm{max}}(t))| \, U''(X_{0,1}(t))} \, e^{-\Delta U_{0,1}(t)}
\, .
\end{equation}
Here, the $X_{0,1}(t)$ denote the minima of the potential wells in state 0 (first subscript) and state 1 (second subscript),
$X_{\mathrm{max}}(t)$ denotes the (smooth) potential maximum between the two states, and
$\Delta U_{0,1}(t)=U(X_{\mathrm{max}}(t)) - U(X_{0,1}(t))$ is the height of the potential barrier to be crossed
when jumping from state 0 (first subscript) to state 1, or from state 1 (second subscript) to state 0.
In the case of a cusp shaped potential barrier, the rate is \cite{hanggi1990}
\begin{equation}
\label{eq:rate_cuspMax}
r_{0,1}(t)=\sqrt{\frac{\Delta U_{0,1}(t)}{\pi}} \, U''(X_{0,1}(t)) \, e^{-\Delta U_{0,1}(t)}
\, .
\end{equation}
Note that the rate expressions Eqs.~\eqref{eq:rate_smoothMax}, \eqref{eq:rate_cuspMax} are dimensionless and are given in units of the inverse time $K/\gamma$.
}

\textcolor{black}{%
The positions of the extrema of the perturbed potential $U(X, t)$ at time $t$ can be formally expressed by the inverse of $V^{\prime}(X)$, that is $\left(V^{\prime}\right)^{-1}$ (cf. Eq. \eqref{eq:Ugeneral}), but in general have to be determined numerically. For the minima,
\begin{subequations}
\begin{equation}
X_{0,1}(t) = \mp a \pm (V')^{-1}(\pm F_\mathrm{erase}(t))
\, ,
\end{equation}
where the upper (lower) signs refer to state 0 (1),
and $(V')^{-1}(F_\mathrm{erase}(0))=(V')^{-1}(0)=0$.
For cusp shaped potentials the inverse $(V')^{-1}$ is usually unique. For potentials with a smooth maximum there are typically two branches. The branch including $X=0$ gives the minima, and that including $X=a$ gives the maximum;
\begin{equation}
X_{\mathrm{max}}(t) = -a + (V')^{-1}(F_\mathrm{erase}(t))
\, ,
\end{equation}
\end{subequations}
with $(V')^{-1}(F_\mathrm{erase}(0))=(V')^{-1}(0)=a$.
}

\section{Double-parabolic potential and linear erasure protocol}
\label{sec:F(t).lin}
We now apply our general framework to
\textcolor{black}{%
bit erasure in the double-parabolic potential from Eq. \eqref{eq:U} using the linear erasure protocol described in Eq.~\eqref{eq:F(t).lin}.
}%

\subsection{Calculating the work distribution}

\subsubsection{The average and variance of the work}
\label{sec:F(t).lin:W}

The sum $\sum_{i=1}^n (-1)^{i+n} F_{\mathrm{erase}}(t_i)$ appears in both the average work, Eq.~\eqref{eq:<W>}, and its variance, Eq.~\eqref{eq:var(W)}.  Upon substitution of Eq.~ \eqref{eq:F(t)erase.lin} into the sum of we find $\lambda \sum_{i=1}^n (-1)^{i+n} t_i$.  The times at which a particle jumps from the left well (state 0) to the right well (state 1) are denoted with a positive sign, while the times at which a particle jumps from state 1
to state 0 are denoted with a negative sign.  (Recall that $n$ is odd whenever a particle is initially,  $t=t_0=0$, located in the left well and even when it is initially located in the right well).  Therefore, the sum gives  the \emph{total} amount of time a particle spends in state 0 during the first phase $[0,\tau^*]$ of the erasure protocol.  Denoting this time by $\tau_0$, we obtain from Eq. \eqref{eq:<W>} the following simple expression for the total average work required to erase one bit of information with the linear protocol \eqref{eq:F(t).lin}; 
\begin{equation}
\label{eq:workRes}
\langle W \rangle = 2\lambda a \langle \tau_0 \rangle
= 2a^2 \frac{\langle \tau_0 \rangle}{\tau^*}
\end{equation}
where we used $\lambda = a/\tau^*$.

Similarly, the variance  of Eq.~\eqref{eq:var(W)} reduces ostensibly to the variance of the residence time $\tau_0$,
\begin{equation}
\label{eq:varWRes}
\Var(W) = a^2 \left( \frac{1}{\tau^*} + \frac{1}{\tau-\tau^*} \right)
	+ 4 a^4 \frac{\langle \tau_0^2 \rangle - \langle \tau_0 \rangle^2}{(\tau^*)^2}
\, ,
\end{equation}
wherein the additional contribution appearing in the first term represents the  accumulated variance from the fluctuating motion within the potential wells.  This contribution arises from the first term in Eq.~\eqref{eq:var(W)} by using the fact that $\dot{F}(t)$ is constant except for a jump at $t=\tau^*$.  Since $\langle \tau_0^2 \rangle - \langle \tau_0 \rangle^2$ is typically of the same order as $(\tau^*)^2$, and since $a/\tau^* = \lambda \ll 1$
the first term in Eq.~\eqref{eq:varWRes} is much smaller than the second term.  Therefore, the fluctuations within the potential wells are negligible and the variance of the work is dominated by the distribution of the sojourn time $\tau_0$ in state 0,

For completeness and convenience, we write Eqs.~\eqref{eq:workRes} and \eqref{eq:varWRes} for the average work and its variance in dimensional form; 
\begin{subequations}
\begin{align}
\label{eq:workDim}
\langle W \rangle
& = 2 K a^2 \frac{\langle \tau_0 \rangle}{\tau^*} \qquad \text{and}
\\[1ex]
\label{eq:varWDim}
\Var(W)
& = a^2 \kB T \gamma \left( \frac{1}{\tau^*} + \frac{1}{\tau-\tau^*} \right)
\nonumber \\
& \quad\quad \mbox{}
	+ \left(2Ka^2\right)^2 \frac{\langle \tau_0^2 \rangle - \langle \tau_0 \rangle^2}{(\tau^*)^2}
\, .
\end{align}
\end{subequations}

Finally, we emphasize that for the averages $\langle \tau_0 \rangle$ and $\langle \tau_0^2 \rangle$, only the total time that a particle spends in state 0 is relevant, and no other details of the trajectory $X(t)$ are necessary.  Therefore, we can write these averages solely using the probability density function\ $\mathcal{P}(\tau_0)$ for $\tau_0$; 
\begin{equation}
\label{eq:<tau0k>}
\langle \tau_0^k \rangle = \int_0^{\tau^*} \mathrm{d}\tau_0 \, \mathcal{P}(\tau_0) \tau_0^k
\qquad (k>0)
\, ,
\end{equation}
and hence we need not consider the full jump-time statistics treated in $\mathcal{P}_n(t_1,t_2,\ldots,t_n)$, allowing us to simplify the notation by suppressing any subscript $\{t_i\}$.  Next, we show how to specialize the approach described in \S~\ref{sec:P(t_i)} to $\mathcal{P}(\tau_0)$. In particular, starting from $\mathcal{P}_n(t_1,t_2,\ldots,t_n)$ we express $\mathcal{P}(\tau_0)$ in terms of the
probabilities $P_{0 \rightleftarrows 1}(\underline{t},\overline{t})$ from Eq.~\eqref{eq:PS}.

\subsubsection{Evaluation of $\mathbf{\mathcal{P}(\tau_0)}$}
\label{sec:methods}

Equipped with Eqs \eqref{eq:workRes} and \eqref{eq:varWRes}, we have reduced the problem of calculating the average work and its variance to that of finding the probability density function $\mathcal{P}(\tau_0)$ for the total residence time of a particle in state 0.  The complete statistics of jump times is contained in the functions $\mathcal{P}_n(t_1,t_2,\ldots,t_n)$ for any number of jumps $n>0$.  Therefore, we derive $\mathcal{P}(\tau_0)$ from $\mathcal{P}_n(t_1,t_2,\ldots,t_n)$ by demanding that $(t_n-t_{n-1}) + (t_{n-2}-t_{n-3}) + \ldots = \tau_0$ and hence constraining the total residence time $\tau_0$ in state 0,
\begin{widetext}
\begin{align}
\label{eq:P0def}
\mathcal{P}(\tau_0) 
& = \left\langle \sum_{n=1}^\infty \delta(\tau_0 - t_n + t_{n-1} - \ldots \mp t_1) \right\rangle_{\!\!\{t_i\}}
\nonumber \\
& = \sum_{n=1}^\infty \int_0^{\tau^*} \mathrm{d}t_1 \int_{t_1}^{\tau^*} \mathrm{d}t_2 \, \ldots \int_{t_{n-1}}^{\tau^*} \mathrm{d}t_n \,
		\delta(\tau_0 - t_n + t_{n-1} - \ldots \mp t_1) \mathcal{P}_n(t_1,t_2,\ldots,t_n)
\, ,
\end{align}
\end{widetext}
where the minus (plus) sign at $\mp t_1$ in the delta function refers to $n$ odd (even).   Note that this distribution $\mathcal{P}(\tau_0)$ does not contain the contributions for $\tau_0=0$, because in Eq.~\eqref{eq:P0def} we assume that at least one jump will occur. Importantly, since $\tau_0=0$ does not contribute to averages of the form \eqref{eq:<tau0k>} we can incorporate it by simply adding $\tfrac{1}{2}\delta(\tau_0)S_1(0,\tau^*)$ to Eq.~\eqref{eq:P0def} (see also Fig.~\ref{Fig4}).

Using Eq.~\eqref{eq:P(t_i)} for $\mathcal{P}_n(t_1,t_2,\ldots,t_n)$ in Eq.~\eqref{eq:P0def}, we can write $\mathcal{P}(\tau_0)$ as a weighted average over all possible jumps,
\begin{equation}
\label{inst_pdf}
\mathcal{P}(\tau_0) = \sum_{n=1}^\infty \Pi_n \mathcal{R}_n(\tau_0)
\, .
\end{equation}
Here, $\mathcal{R}_n(\tau_0)$ is the probability of observing $n$ jumps during the time interval $[0,\tau^*]$ at times $0 \leq t_1 < t_2 < \ldots < t_n \leq \tau^*$ amounting to a total residence time $\tau_0$ in state 0,
\emph{relative} to the probability of observing $n$ jumps at any arbitrary sequence of times $0 \leq t_1 < t_2 < \ldots < t_n \leq \tau^*$,
\begin{widetext}
\begin{align}
\label{eq:Rn}
\mathcal{R}_n(\tau_0) & =
\frac{\int_0^{\tau^*}\mathrm{d}t_1 P_{0 \rightleftarrows 1}(0,t_1)
	\int_{t_1}^{\tau^*}\mathrm{d}t_2 P_{0 \leftrightarrows 1}(t_1,t_2) \, \cdots
	P_{0\rightarrow1}(t_{n-1},\tau_0 + t_{n-1} - \ldots \mp t_1)S_1(\tau_0 + t_{n-1} - \ldots \mp t_1,\tau^*)}
	{\int_0^{\tau^*}\mathrm{d}t_1 P_{0 \rightleftarrows 1}(0,t_1)
	\int_{t_1}^{\tau^*}\mathrm{d}t_2 P_{0 \leftrightarrows 1}(t_1,t_2) \, \cdots
	\int_{t_{n-1}}^{\tau^*}\mathrm{d}t_n P_{0\rightarrow1}(t_{n-1},t_n)S_1(t_n,\tau^*)}
\, ,
\end{align}
\end{widetext}
where, as in Eq.~\eqref{eq:PIn}, the upper (lower) arrows in the subscripts refer to $n$ odd (even).

\textcolor{black}{%
In terms of the rates $r_{0,1}(t)$, the jump probabilities $P_{0 \rightleftarrows 1}(\underline{t},\overline{t})$ and $S_{0,1}(\underline{t},\overline{t})$ are defined in Eqs.~\eqref{eq:PS} and \eqref{eq:S01} respectively. For the double-parabolic potential \eqref{eq:U} with a cusp-shaped maximum that is perturbed using the linear erasure protocol \eqref{eq:F(t).lin}, these rates follow from Eq.~\eqref{eq:rate_cuspMax} as (see also \cite{survival,SR})
\begin{equation}
\label{eq:P01.lin}
\begin{split}
& P_{0 \rightleftarrows 1}(\underline{t},\overline{t})
= \frac{a(1 \mp \overline{t}/\tau^*)}{2\sqrt{\pi}}  e^{-\frac{1}{2}a^2(1 \mp \overline{t}/\tau^*)^2}
\\
& \mbox{}\times
\exp\left[
	-\frac{\tau^*}{\sqrt{2\pi}a}  e^{-\frac{1}{2}a^2(1 \mp \overline{t}/\tau^*)^2}
		+ \frac{\tau^*}{\sqrt{2\pi}a}  e^{-\frac{1}{2}a^2(1 \mp \underline{t}/\tau^*)^2} 
\right]
\, ,
\end{split}
\end{equation}
where the upper (lower) sign on the right-hand side refers to the upper (lower) arrow in the subscript of $P_{0 \rightleftarrows 1}$.
}
\begin{figure*}
\begin{center}
\includegraphics[width=0.8\linewidth]{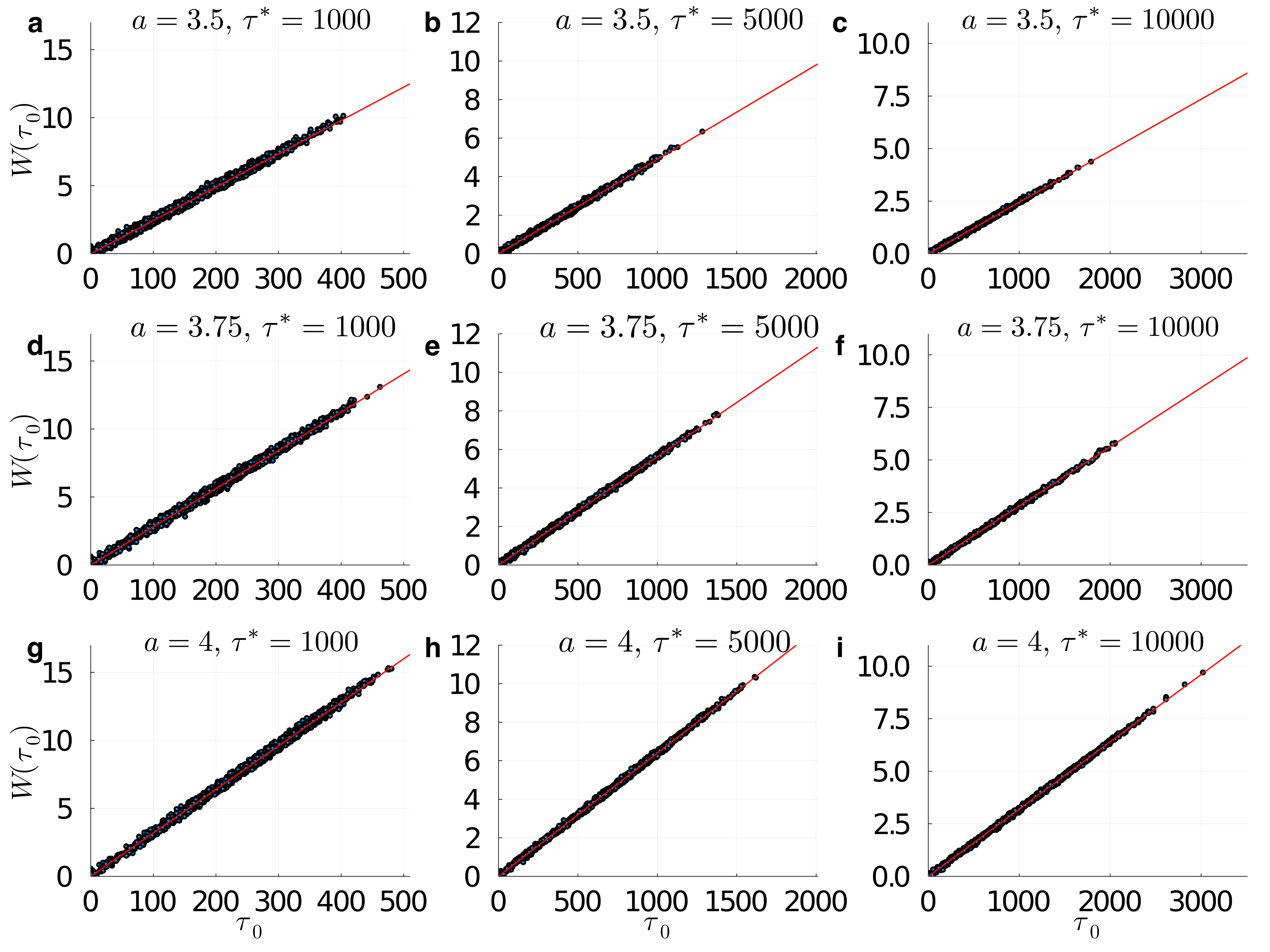}
\caption{Total work for 9600 different trajectories $X(t)$ as a function of $\tau_0$ evaluated from numerical simulations of the Langevin equation \eqref{eq:LE} and equation \eqref{work} without taking the ensemble average $\langle \cdot \rangle$ to obtain the trajectory-based work \cite{Sekimoto} (dots).  This is compared to the analytical prediction $W=2\lambda a \tau_0=2 a^2 \tau_0/\tau^*$, namely the work before taking the average over
the distribution of $\tau_0$ in~\eqref{eq:workRes} (solid red line).  We use $a=3.5$ for (a,b,c); $a=3.75$ for (d,e,f); $a=4$ for (g,h,i);  $\tau^*=1000$ for (a,d,g); $\tau^*=5000$ for (b,e,h); $\tau^*=10000$ for (c,f,i).}
\label{Fig3}
\end{center}
\end{figure*}
\begin{figure*}
\begin{center}
\includegraphics[width=0.9\linewidth]{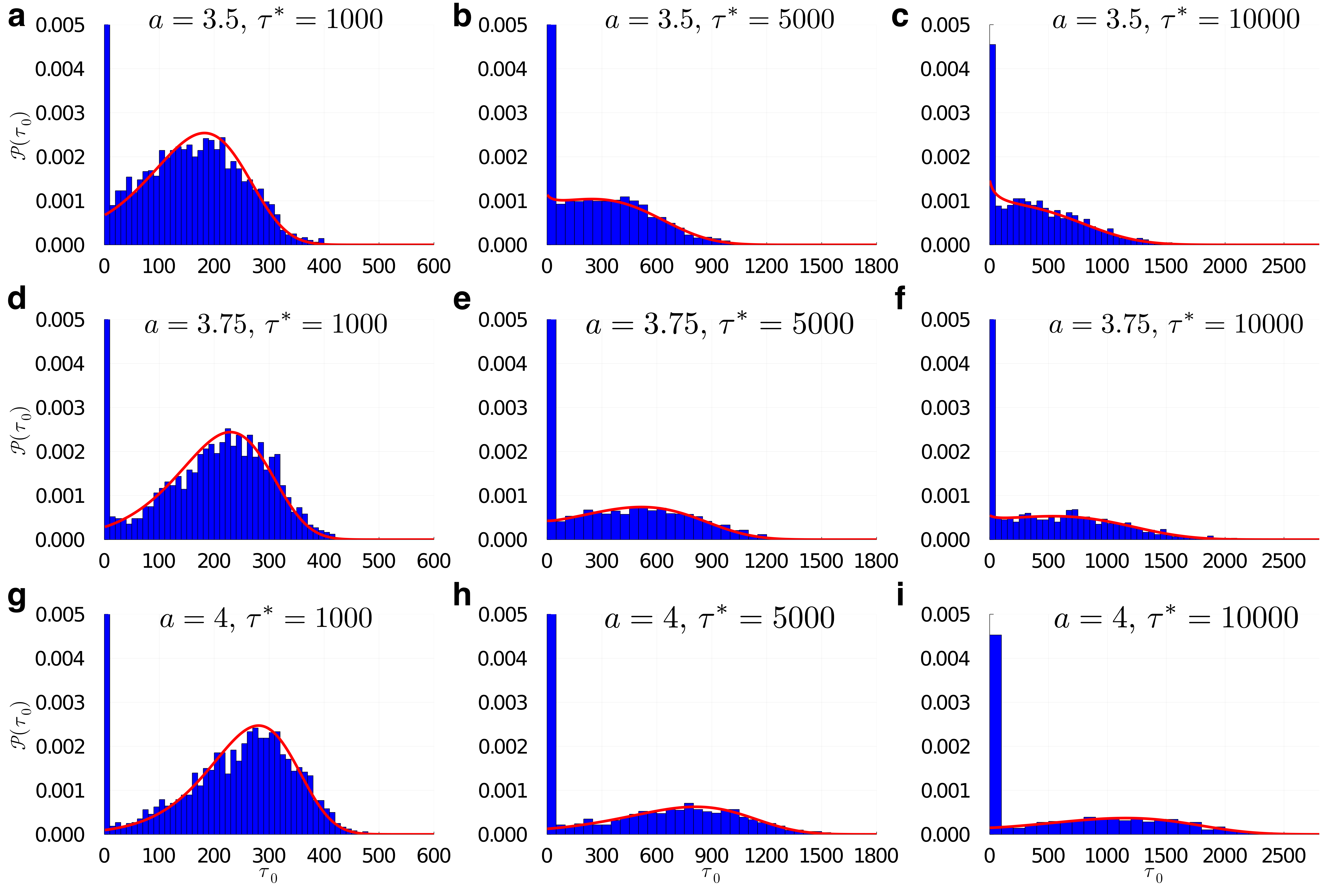}
\caption{Comparison of $\mathcal{P}(\tau_0)$ obtained by direct numerical simulations (blue histogram) to our analytical prediction Eq.~\eqref{inst_pdf} with three jumps (Eq. \eqref{eq:N3}) (solid red line) \cite{note1}.
We did not include the contribution $n=0$ of zero transitions in theoretical curves from \eqref{inst_pdf} plotted here, which are thus restricted to positive times $\tau_0$. Note that this contribution is irrelevant for averages of the form $\langle \tau_0^k \rangle$ ($k>0$) (see also the discussion following Eq.~\eqref{eq:P0def}).
The parameter values are the same as in Fig.~\ref{Fig3}: $a=3.5$ for (a,b,c); $a=3.75$ for (d,e,f); $a=4$ for (g,h,i);  $\tau^*=1000$ for (a,d,g); $\tau^*=5000$ for (b,e,h); $\tau^*=10000$ for (c,f,i).}
\label{Fig4}
\end{center}
\end{figure*}

\subsection{Analysis of the work distribution}
\label{sec:results}

\subsubsection{The work as a function of sojourn time $\tau_0$}

We now examine Eqs.~\eqref{eq:workRes} and \eqref{eq:varWRes} for the average work and its variance associated with erasing one bit of information through the linear protocol Eq.~\eqref{eq:F(t).lin}.
In Eq.~\eqref{eq:workRes}, the complete average over the thermal fluctuations and the associated ensemble of trajectories is reduced to an average $\langle \tau_0 \rangle$ of the time $\tau_0$ that a particle spends in state 0 during the erasure process. The variance of the work, Eq.~\eqref{eq:varWRes}, is a combination of the variance of $\tau_0$ and the accumulated fluctuations that the trajectories experience near the potential well minima in state 0 and in state 1.  Therefore, we predict that for all trajectories with the same $\tau_0$ the work depends linearly on $\tau_0$, where the variations about the value $2\lambda a \tau_0$ stem from the
fluctuations within the potential wells.

We test this prediction with numerical simulations of the Langevin equation \eqref{lang} for many independent realizations of the erasure process.  First, we calculate the work for each numerical trajectory $X(t)$ by evaluating the right-hand side of Eq.~\eqref{eq:W}. Second, we sort the resulting work values according to the time $\tau_0$ a trajectory spent in state 0.  In Fig.~\ref{Fig3} we compare the analytical prediction, $2\lambda a \tau_0=2 a^2 \tau_0/\tau^*$, with the numerical results for the work as a function of $\tau_0$, for various values of $\tau^*$ and $a$.  The comparison is splendid, with only minor fluctuations around this linear behavior, the origin of which is the relative magnitude of the two summands in Eq.~\eqref{eq:varWRes} as discussed above. Namely, the fluctuations associated with the noise-driven transitions between the two states 0 and 1, which are captured in the distribution $\mathcal{P}(\tau_0)$ of sojourn times $\tau_0$, dominates the thermal fluctuations of the trajectories within the potential wells.

\subsubsection{Distribution of sojourn times $\tau_0$}

We now test our analytical prediction for the distribution $\mathcal{P}(\tau_0)$ of the sojourn times $\tau_0$, as given in Eqs.~\eqref{inst_pdf} and \eqref{eq:Rn}.  We use the approximation \eqref{eq:N3} for three transitions, which is the most accurate of the alternatives discussed in \S~\ref{sec:nfinite}.  Thus, to determine $\mathcal{P}(\tau_0)$ from Eq.~\eqref{inst_pdf} we truncate the sum at $N=3$ transitions and then evaluate the integrals in $\Pi_n$ and $\mathcal{R}_n(\tau_0)$, with $n \leq 3$, by numerical integration (c.f. \eqref{eq:N3} and \eqref{eq:Rn}).  In Fig.~\ref{Fig4} we again compare the simulations of the Langevin equation \eqref{lang}, for many independent realizations of the erasure process, to our analytical prediction for $\mathcal{P}(\tau_0)$, for the same parameter values as in Fig.~\ref{Fig3} \cite{note1}.

The analytical results agree extremely well with the simulations.  The deviation between the simulation histograms and the analytical results at $\tau_0=0$ is a consequence of excluding the $n=0$ term in the sum \eqref{inst_pdf} of the analytic prediction. Because the corresponding trajectories start and end in state 1 without ever jumping to state 0, they do not contribute to the average work \eqref{eq:workRes}, and we excluded them in $\mathcal{P}(\tau_0)$ as shown in Fig.~\ref{Fig4}.

The dependence of $\mathcal{P}(\tau_0)$ on the values of $a$ and $\tau^*$ is evident in Fig.~\ref{Fig4}.  For small times $\tau^*$, $\mathcal{P}(\tau_0)$ is peaked about intermediate values of the sojourn time $\tau_0$ (leftmost column in Fig.~\ref{Fig4}).  It is intuitive that as the protocol duration decreases, the largest contribution to $\mathcal{P}(\tau_0)$ comes from one-jump trajectories. The probability of a transition from state 0 to 1
increases rapidly due to the rapid tilting of the potential, thereby decreasing the survival probability in state 0 accordingly.  Clearly, since the contribution to $\mathcal{P}(\tau_0)$ from one-jump trajectories is the product of  these rapidly increasing and decreasing quantities, we observe the $\tau_0$ and $\tau^*$ dependent peaks in $\mathcal{P}(\tau_0)$.  As $\tau^*$ increases (middle and right columns of Fig.~\ref{Fig4}),
the two minima of the potential have similar energies for a longer time period at the beginning of the erasure protocol, during which the probabilities of the system jumping from state 0 to state 1 and {\em vice-versa}
remain comparable. This feature increases the relative weight of a larger number of inter-well transitions and is the origin of the spread of $\mathcal{P}(\tau_0)$ for longer protocol duration.

Finally, increasing $a$ (the three rows in Fig.~\ref{Fig4}) increases the height of the potential barrier between the wells at $X=0$.   Therefore, a longer time is required to tilt the potential sufficiently far to drive
transitions from state 0 into state 1, thereby shifting the characteristic features of $\mathcal{P}(\tau_0)$ towards larger values of $\tau_0$.
\begin{figure*}
\begin{center}
\includegraphics[width=.8\linewidth]{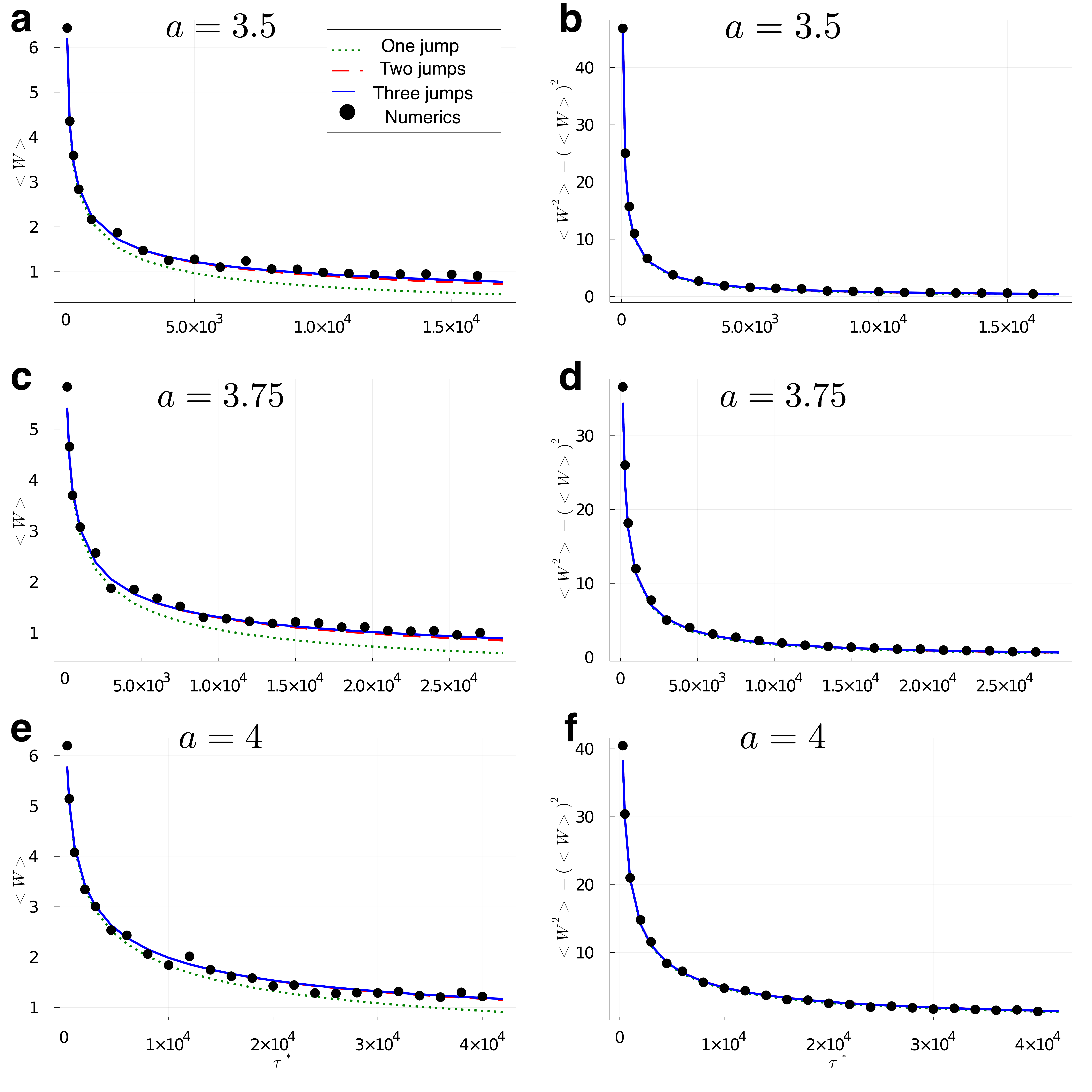}
\caption{Total average work, $\langle W \rangle$ (a,c,e), and its variance, $\langle W^2 \rangle - \langle W \rangle^2$ (b,d,f), for different durations of the erasure protocol obtained from numerical simulation of the erasure process (black dots) and analytically from Eqs.~\eqref{eq:workRes} and \eqref{eq:varWRes} \cite{note1}, with one (green dotted lines, \eqref{eq:N1}), two (red dashed lines, \eqref{eq:N2}) and three (blue solid lines, \eqref{eq:N3}) transitions between the two states.  Clearly, the approximations for these three transitions are indistinguishable in the variance (b,d,f).
Here we use: $a=3.5$ for (a, b), $a=3.75$ for (c,d), $a=4$ for (e,f).}
\label{Fig5}
\end{center}
\end{figure*}

\subsubsection{Erasing bits in finite-time}

As seen in Eq.~\eqref{inst_pdf}, an exact calculation of $\mathcal{P}(\tau_0)$ involves an infinite number of terms; summing the contributions from any number $n$ of transitions.  However, the results of Fig.~\ref{Fig4}
show that we need only consider three jumps between the two potential wells in order to obtain an accurate estimate of $\mathcal{P}(\tau_0)$.  We now analyze the influence of the number of jumps on the average work  and its variance by comparing the cases of one, two and three transitions; Eqs.~\eqref{eq:N1}, \eqref{eq:N2} and \eqref{eq:N3}. In Fig.~\ref{Fig5} we show $\langle W \rangle$ and $\Var(W)$ as a function of $\tau^*$, as predicted for these three cases from Eqs. \eqref{eq:workRes} and \eqref{eq:varWRes} (solid lines). We also show the results from direct numerical simulations of the Langevin equation \eqref{lang} (dots), for which we calculated the work using Eq. \eqref{work}, with no restriction on the number of jumps. 

Firstly, as expected, both $\langle W \rangle$ and $\Var(W)$ are decreasing functions of $\tau^*$ and, consistent with previous findings \cite{aure1,berut,berut1,bech,bech1,dago,dago1}, the average work approaches the lower Landauer bound only for quasi-statically slow erasure processes.  Secondly, we see the striking energetic costs--increased work or thermal dissipation--as information is erased more rapidly.   Thirdly, very few jumps are required to give extremely accurate predictions of the average work for rapid erasure processes. For example, for $N=1$ jump between state 0 and 1, our theory provides very accurate estimates for $\tau^* \lesssim 1000$, which is improved over a longer range of times for $N=2$ jumps, ostensibly saturating for $N=3$ jumps.  These results are consistent with our intuition that three-jump processes ($0\rightarrow1\rightarrow0\rightarrow1$) are  significantly less likely to occur than one- or two-jump processes.  Finally, for $\Var(W)$ all of the predictions are indistinguishable.  

The analytical predictions will become less accurate as $\tau^*$ increases and the number of inter-well transitions increases.  In particular, any finite-transition approximation will underestimate the average work,
because trajectories with many transitions between the two states increase the total average time $\langle \tau_0 \rangle$ the particle spends in state 0, as seen in Eq.~\eqref{eq:workRes}. 
Therefore, in principle, accurate reproduction of the Landauer bound \cite{Landauer1961,land1988} from Eqs.~\eqref{eq:workRes} and \eqref{inst_pdf} should treat infinitely many transitions for a quasi-statically slow ($\tau^* \to \infty$) erasure protocol.

\subsubsection{The Landauer bound}

Now we derive the Landauer bound from Eq.~\eqref{eq:workRes} without explicitly evaluating the average $\langle \tau_0 \rangle$. To display the role of the thermal energy, $\kB T$, we use dimensional quantities.  
As noted throughout, the Landauer bound \cite{Landauer1961,land1988} is reached for quasi-statically slow erasure protocols; $\lambda \to 0$ and $\tau^* \to \infty$, with $\lambda \tau^* = K a$ fixed to achieve perfect erasure.  The slow erasure protocol implies that the particle distribution at any time $t \in [0,\tau^*]$ is governed by the momentarily equilibrium Boltzmann distribution, 
\begin{subequations}
\begin{equation}
p_{\mathrm{qs}}(X) = \frac{1}{Z_{\mathrm{qs}}} \exp \left[-\frac{U_{\mathrm{qs}}(X(t))}{\kB T} \right] 
\, ,
%
\end{equation}
with potential $U_{\mathrm{qs}}(X(t)) \equiv \tfrac{K}{2}[X(t) -\textrm{sign}(X(t)) a]^2  - \lambda t X$, and normalization
\begin{multline}
Z_{\mathrm{qs}} =
\sqrt{\frac{\pi \kB T}{2 K}} e^{\frac{\lambda t (\lambda t - 2Ka)}{2 K \kB T}}
	\left( 1 + \mathrm{erf}\left[ \frac{Ka - \lambda t}{\sqrt{2 K \kB T}} \right]
	\right)
\\ +
\sqrt{\frac{\pi \kB T}{2 K}} e^{\frac{\lambda t (\lambda t + 2Ka)}{2 K \kB T}}
	\left( 1 + \mathrm{erf}\left[ \frac{Ka + \lambda t}{\sqrt{2 K \kB T}} \right]
	\right)
\, ,
\end{multline}
\end{subequations}
where $\mathrm{erf}(x) = \frac{2}{\sqrt{\pi}}\int_0^x \mathrm{d}y \, e^{-y^2}$ is the error function.

Under such equilibrium conditions the ratio of state 0 to state 1 sojourn times is identical to the ratio of probability weights within the corresponding potential wells. Let $\mathrm{d}t$ be the time increment around a specific 
time $t$ during the erasure protocol.  Hence, the fraction of time spent in state 0 is $\mathrm{d}t \int_{-\infty}^0 \mathrm{d}X \, p_{\mathrm{qs}}(X)$ and we find that the average time a particle spends in state 0, relative to the duration $\tau^*$ of the full erasure procedure, is
\begin{align}
\label{eq:tau0_qs}
\frac{\langle \tau_0 \rangle}{\tau^*}
& = \int_0^{\tau^*} \frac{\mathrm{d}t}{\tau^*} \int_{-\infty}^0 \mathrm{d}X \, p_{\mathrm{qs}}(X)
\nonumber \\
& = \int_0^{\tau^*} \frac{\mathrm{d}t}{\tau^*} \left\{
	\left(1 + e^{\frac{2a\lambda t}{\kB T}}\right) \frac{1 + \mathrm{erf}\left[ \frac{Ka + \lambda t}{\sqrt{2 K \kB T}} \right]}
										   {1 + \mathrm{erf}\left[ \frac{Ka - \lambda t}{\sqrt{2 K \kB T}} \right]}
	\right\}^{\!\!-1}
	\, .
\end{align}

Evaluation of Eq.~\eqref{eq:tau0_qs} appeals to the theory underlying our main result Eq.~\eqref{eq:workRes}, which is valid when the potential barrier between the two memory states is much larger than the thermal energy (see \S~\ref{sec:model}).  We examine this constraint for quasi-static information erasure by first observing the consequence of $\kB T \ll K a^2$ in Eq.~\eqref{eq:tau0_qs}.  The poorly converging series representation of $\mathrm{erf}(x)$ precludes a useful formal asymptotic expansion, but because $Ka > \lambda t$, we take $\mathrm{erf}(x)$ = 1 for $x\gg$1, so that Eq.~\eqref{eq:tau0_qs} becomes
\begin{align}
\frac{\langle \tau_0 \rangle}{\tau^*}
& \approx \frac{1}{\tau^*}\int_0^{\tau^*} \frac{\mathrm{d}t}{1 + e^{\frac{2a\lambda t}{\kB T}}}
\nonumber \\
& = 1 + \frac{\kB T \ln 2 - \kB T \ln \left[ 1 + e^{\frac{2a\lambda\tau^*}{\kB T}} \right]}{2a\lambda\tau^*}
\nonumber \\
& \approx \frac{\kB T \ln 2}{2a\lambda\tau^*}
\, ,
\end{align}
where the final expression results from $\lambda \ll 1$, $\tau^* \gg 1$, with $\lambda\tau^* = Ka$ fixed.  Substitution in Eq.~\eqref{eq:workDim} with $\lambda \tau^* = K a$ yields the Landauer bound, 
\begin{equation}
\langle W \rangle = \kB T \ln 2,
\end{equation}
which is the minimal amount of work required, on average, to quasi-statically erase one bit of information or, equivalently, the least average amount of heat dissipated into the thermal 
environment \cite{note2}.

\begin{figure*}
\begin{center}
\includegraphics[width=.8\linewidth]{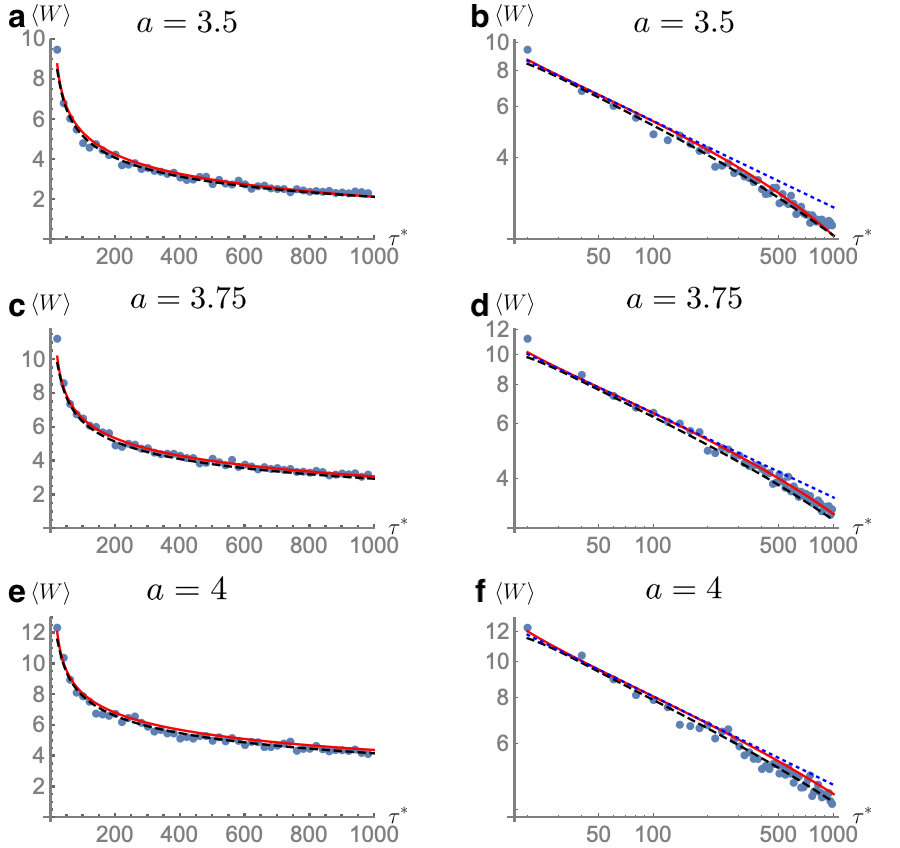}
\caption{Mean work obtained from numerical simulations of the Langevin equation \eqref{lang} (blue dots), compared to the analytical approximation Eq.~\eqref{W_appr1} (solid red lines), the improved approximation
obtained by replacing $P_{0 \rightarrow 1}(0,t)$ with $g(t)$ from Appendix \ref{app:shortProtocol} in Eq.~\eqref{eq:meanW} (black dashed lines), and the power-law approximation Eq.~\eqref{W_appr2} (blue short-dashed lines) \cite{note1}.
In the left (right) column we use a linear (log-log) scale, with $a=3.5$ (a, b), $a=3.75$ (c, d), and $a=4$ (e, f).}
\label{Fig6}
\end{center}
\end{figure*}
\subsubsection{Erasing bits rapidly}

\label{sec:fast}

Here we consider the rapid erasure of information; the limit opposite to that of Landauer bound.   In our framework, the fastest possible linear erasure protocol is realized when $F_{\mathrm{erasure}}(t)$ is changing so
rapidly that the probability a particle will jump from state 1 to state 0 within the erasure phase $t \in [0,\tau^*]$ is negligibly small, even though $\dot{F}_{\mathrm{erasure}}(t) \ll 1$.  If a particle is in state 0 at the beginning of the erasure process it will jump only once to state 1, a process accurately described by Eq.~\eqref{eq:N1} for one inter-well transition.  See also Fig.~\ref{Fig5}, where we demonstrate that the $N=1$ approximation provides excellent accuracy for fast erasure protocols.  Therefore, we restrict the sum in \eqref{inst_pdf} to one transition and use the explicit expression for $\mathcal{R}_1(\tau_0)$ to write the average work \eqref{eq:workRes} as
\begin{align}
\label{eq:meanW}
\langle W \rangle &\simeq
\frac{a^2}{\tau^*}
\frac{\int_0^{\tau^*}\mathrm{d}\tau_0 \, P_{0 \rightarrow 1}(0,\tau_0) \tau_0}
	 {\int_0^{\tau^*}\mathrm{d}t \, P_{0 \rightarrow 1}(0,t)}
\, ,
\end{align}
where $S_1(\tau_0,\tau^*) \simeq 1$ for all $\tau_0 \geq 0$ in $\mathcal{R}_1(\tau_0)$.  This is the same approximation of $S_1(0,\tau^*)$ leading to the probabilities of zero or one inter-well transition in Eq.~\eqref{eq:N1}.

The blue curves in Fig.~\ref{supp_Fig1} show that for smaller values of $\tau^*$, or faster erasure protocol, $P_{0 \rightarrow 1}(0,\tau_0)$ becomes more symmetric about the maximum at $\tau_{\mathrm{max}}$, ostensibly independent of $a$.  Were $P_{0 \rightarrow 1}(0,\tau_0)$ perfectly mirror-symmetric about $\tau_0=\tau_{\mathrm{max}}$, then 
$\tfrac{\int_0^{\tau^*}\mathrm{d}\tau_0 \, P_{0 \rightarrow 1}(0,\tau_0) \tau_0}{\int_0^{\tau^*}\mathrm{d}t \, P_{0 \rightarrow 1}(0,t)} = \tau_{\mathrm{max}}$ exactly.
We now assume mirror-symmetry to calculate the average work in Eq.~\eqref{eq:meanW} \textcolor{black}{with $P_{0 \rightarrow 1}(0,\tau_0)$ given in Eq.~\eqref{eq:P01.lin}}.  Using the linear protocol of Eq.~\eqref{eq:F(t).lin} we obtain
\begin{equation}
\label{eq:tauMax}
\tau_{\mathrm{max}}=\tau^* \left( 1 - \tfrac{1}{a}\sqrt{2 \ln \left[\tfrac{\tau^*}{a\sqrt{2\pi}}\right]} \right)
\end{equation}
as the value of $\tau_0$ where $P_{0 \rightarrow 1}(0,\tau_0)$ is a maximum.  Therefore, we find that the average work for rapid erasure protocols is approximately
\begin{equation}
\label{W_appr1}
\langle W \rangle \simeq a^2-a\sqrt{2 \ln \left[\frac{\tau^*}{a\sqrt{2\pi}}\right]}
\, .
\end{equation}
Although this is not an asymptotically exact approximation, in the sense that is not obtained from a systematic expansion procedure, Fig.~\ref{Fig6} shows how well it reproduces the numerical results for $\langle W \rangle$ for rapid-erasure conditions.  Thus, it provides a simple functional form of the average work for small $\tau^*$.  In Appendix \ref{app:shortProtocol} we describe a more accurate approximation, which, although significantly more complicated, also allows estimation of the variance of the work.  Figs.~\ref{Fig6} and \ref{supp_Fig3} show that this improved approximation is nearly identical to the numerical simulation results.

The right (log-log) column of Fig.~\ref{Fig6} shows a roughly linear $\langle W \rangle$ versus  $\tau^*$ behavior.  Extracting the corresponding exponent from Eq.~\eqref{W_appr1}, we find that
for rapid erasure protocol we can further approximate $\langle W \rangle$ as
\begin{subequations} 
\begin{equation}
\label{W_appr2}
\langle W \rangle \simeq C(a)\, \left(\tau^*\right)^{-\frac{4}{a^2}}
\, ,
\end{equation}
with
\begin{equation}
C(a) = \frac{a^2}{2} \sqrt{e}\left( 2\pi a^2 \right)^{\frac{2}{a^2}}
\, .
\end{equation}
\end{subequations}
Rewriting \eqref{W_appr2} in dimensional form,
\begin{equation}
\langle W \rangle \simeq C(a)\, \left(\tau^*\right)^{-\frac{4 \kB T}{K a^2}}
\, ,
\end{equation}
we find that the exponent is twice the inverse of the height of the potential barrier between the two memory states, measured in units of the thermal energy $\kB T$.   Therefore, as $\tau^*$ decreases and erasure becomes more rapid, the average work increases with a rate set by the barrier height.
Since our fast-relaxation assumption requires $\tau^* \gg 1$, this dependence on $\tau^*$ is significantly weaker than the known
optimal bound, $\propto 1/\tau^*$, obtained when minimizing the dissipation for non-quasistatic transitions of duration $\tau^*$ between two system states \cite{aure1,zulk,zulk1,boyd1,bech,bech1,zhang,zhang1}.
For information erasure this provides the optimal approach to the Landauer limit as $\tau^* \to \infty$ \cite{aure1,zulk,zulk1,bech,bech1}.
However, optimal erasure processes typically require complicated deformations of the double-well potential with steep gradients or even discontinuities \cite{aure1,zulk,zulk1,bech,bech1}.  In contrast, our result, Eq.~\eqref{W_appr2}, is valid for the non-optimized, strictly linear erasure protocol of Eq.~\eqref{eq:F(t).lin}.  Note that the exponent in Eq.~\eqref{W_appr2} becomes $-1$ for $a=2$ (or dimensionally, $Ka^2=4\kB T$),
and therefore cannot be valid for smaller values of $a$.  This cut-off is consistent with the high potential barrier between the memory states, which is guaranteed by $a^2 \gg 1$ (see \S~\ref{sec:model}).

\section{Conclusions}
\label{sec:conclusions}

We have examined the statistical thermodynamics of erasing a classical bit of information when thermal fluctuations play a significant role.  Our general framework is the common representation of bits of information in small systems \cite{parr,aure1,bech,bech1,berut,berut1,dago,dago1}.  The bits of information are states--zero or one--in a double well potential, and their trajectories are treated as a Brownian particle governed by a Langevin equation.  Our specific approach allowed us to derive analytical results characterizing the distribution of the work required to erase a classical bit of information in finite time.  The work so derived is the heat 
dissipated into a thermal bath during the erasure process.  For the symmetric memories considered here,  where the initial probabilities of finding the memory in state 0 or 1 are equal, the average work and the average heat dissipated are identical.

A key step in our analytical approach is to connect the work performed during the erasure process to the temporal statistics of particle jumps from one memory state to the other.  In consequence, we derived formulae for the average work (Eq.~\eqref{eq:<W>}) and its variance (Eq.~\eqref{eq:var(W)}), where the averages are taken over the jump-time distribution for arbitrary erasure protocols.  We find that these quantities are dominated by the jump statistics between the potential wells and that fluctuations within the wells play only a minor role.
\textcolor{black}{The jump statistics are generated
from the survival probability within the individual states, for which analytical expressions are given in terms of the escape rates.}  For the case of a linearly increasing erasure force (Eq.~\eqref{eq:F(t).lin}) we showed that the work is proportional to the cumulative time that a particle spends in state 0, while jumping back and forth between the two potential wells, leading to simple proportionalities between the average work and its variance and the average sojourn time in state 0 (Eqs.~\eqref{eq:workRes} and \eqref{eq:varWRes}).  

We find excellent agreement between our theoretical predictions and direct numerical simulations of Brownian particle motion during the erasure process.  Perhaps surprisingly, this agreement holds even when restricting the theoretical analysis to a maximum of three transitions between memory states 0 and 1 rather than summing over an infinite number of transitions.  The numerical analysis confirmed the dominant role of the inter-well jump statistics over the fluctuations within the potential wells in determining the work distribution (cf.\ Figs.~\ref{Fig3} and \ref{Fig4}).  Hence, our theory captures the essential characteristics of particle trajectories.  Finally, we
reproduce the Landauer bound in the appropriate limits of a quasi-statically slow erasure process and an ``infinitely'' high potential barrier between the memory states.

Our central ideas can be generalized.  For example, to (a) asymmetric memories, (b) erasure protocols that merge the two potential minima by shifting them towards their common mirror symmetry point without tilting the potential \cite{dago,dago1}; (c) physical realizations of one-bit memory in underdamped settings \cite{dago,dago1};
\textcolor{black}{(d) rapidly fluctuating potential wells (escape rates may then be calculated using instantons \cite{lehmann2000,precursors}, or related recently developed methods from stochastic resonance and early warning quantifiers \cite{SR,SR1,precursors,chen})};
(e) devising finite-time erasure protocols that minimize the thermodynamic costs of
information erasure. In this latter case, in the context of optimal protocols for fast memory erasure, we may seek to reduce heat generation during computation or minimize the variance of the work required to erase one bit of information.  More speculative settings in which our approach would be useful include (f) assessing the stability of memory states through the statistics of transitions that alter a specific state.  One could then quantify the trade-offs between the reliability and accuracy of memory \cite{boyd1} versus the heat dissipated when manipulating the memory state; (g) transient information loss or gain in black holes will be associated with a rate-dependent heat dissipation or generation.  Tolman \cite{Tolman} showed that in a weak static gravitational field the equilibrium temperature depends on the gravitational potential, $\Psi$, as $T(1+\tfrac{\Psi}{c^2})$, where $c$ is the speed of light.  It has recently been argued that this effect shifts the Landauer bound \cite{plastino}, which may have an influence on the black hole information paradox \cite{hawk}.  However, although the irreversibility of black hole evaporation may be constrained by the transient effects that we have shown here are a requirement of statistical mechanics, the consequences of the conformity of transient thermodynamics and relativistic causality \cite{Israel} will play out in a black-hole-model dependent manner. Clearly, there are many implications of our analytical framework for finite-time information erasure.  We hope that readers will pursue the scientific tendrils discussed here and the many yet to be recognized.

\begin{acknowledgments}
L.T.G., W.M.\ and J.S.W.\ gratefully acknowledge support from the Swedish Research Council (Vetenskapsr{\aa}det) Grant No. 638-2013-9243.
R.E.\ acknowledges funding by the Swedish Research Council (Vetenskapsr{\aa}det)
under Grant No.~2020-05266.
L.T.G.\ and R.E.\  thank Salamb\^{o} Dago for helpful discussions.  Nordita is partially supported by Nordforsk. 
\end{acknowledgments}

\appendix

\section{Analytical approximations for short duration of the linear protocol}
\label{app:shortProtocol}

Here we derive an analytical approximation of the integrals in Eqs.~\eqref{eq:workRes} and \eqref{eq:varWRes} for rapid erasure protocol, which yield corresponding approximations for the mean and the variance of the work respectively.  The brevity of the protocol insures that the contribution to the mean and variance of the work from trajectories with more than one inter-well transition is negligible.   Using the same approximations as in \S~\ref{sec:fast}, the two quantities of interest, $\langle W \rangle$ and $\Var(W)$,  are
\begin{subequations}
\begin{align}
\label{app_meanW}
\langle W \rangle &\simeq
\frac{a^2}{\tau^*}
\frac{\int_0^{\tau^*}\mathrm{d}\tau_0 \, P_{0 \rightarrow 1}(0,\tau_0) \tau_0}
	 {\int_0^{\tau^*}\mathrm{d}t \, P_{0 \rightarrow 1}(0,t)}
\, \qquad \text{and}
\\
\label{app_varianceW}
\langle W^2 \rangle - \langle W \rangle^2 &\simeq
 \frac{2 a^4}{(\tau^*)^2}
 \frac{\int_0^{\tau^*}\mathrm{d}\tau_0 \, P_{0 \rightarrow 1}(0,\tau_0) \tau_0^2}
 	  {\int_0^{\tau^*}\mathrm{d}t \, P_{0 \rightarrow 1}(0,t)}
- \langle W \rangle^2
\, .
\end{align}
\end{subequations}
In \S~\ref{sec:fast} we approximated $\langle W \rangle$, Eq.~\eqref{app_meanW}, by assuming that $P_{0 \rightarrow 1}(0,\tau_0)$ is perfectly symmetric about its maximum value, $\tau_0=\tau_{\mathrm{max}}$, given by Eq.~\eqref{eq:tauMax}.
Here we derive a more accurate analytical approximation by explicitly using the fact that $P_{0 \rightarrow 1}(0,\tau_0)$ is symmetric only for very small $\tau^*$.

Firstly, we note that because $\lim_{\tau_0 \to 0}P_{0 \rightarrow 1}(0,\tau_0)$ is a Gaussian, the moments follow easily. However, for $\tau_0 \to \infty$ \textcolor{black}{all the terms in $P_{0 \rightarrow 1}(0,\tau_0)$ are relevant}, but the analytical moments are confounded by the presence of exponentials of exponentials.  Therefore, in order to facilitate calculation of the moments we approximate the probability $P_{0 \rightarrow 1}(0,\tau_0)$ with the function $g(\tau_0)$, which is a combination of Gaussians, viz.
\begin{align}
\label{eq:A2}
g(\tau_0)&=g_1(\tau_0) \frac{a(1 - \tau_0/\tau^*)}{2\sqrt{\pi}}  e^{-\frac{1}{2}a^2(1 - \tau_0/\tau^*)^2}  \nonumber \\
&+ g_2(\tau_0)\alpha e^{\beta(\tau_0-\tau_{\mathrm{max}})^2}
\, ,
\end{align}
such that
\begin{equation}
\begin{split}
\lim_{\tau_0 \to 0} g(\tau_0) &= \frac{a(1 - \tau_0/\tau^*)}{2\sqrt{\pi}}  e^{-\frac{1}{2}a^2(1 - \tau_0/\tau^*)^2}  \qquad \text{and}
\\ 
\lim_{\tau_0 \to \tau_{\mathrm{max}}} g(\tau_0) &= \alpha e^{\beta(\tau_0-\tau_{\mathrm{max}})^2}
\, ,
\end{split}
\end{equation}
requiring that $g_1(\tau_0)$ and $g_2(\tau_0)$ satisfy the following four constraints;
\begin{equation}
\begin{split}
\label{g_eqs}
\lim _{\tau_0 \to 0} g_1(\tau_0) = 1
\, , \quad
\lim _{\tau_0 \to \tau_{\mathrm{max}}} g_1(\tau_0) = 0
\, ,
\\
\lim _{\tau_0 \to 0} g_2(\tau_0) = 0
\, , \quad
\lim _{\tau_0 \to \tau_{\mathrm{max}}} g_2(\tau_0) = 1
\, .
\end{split}
\end{equation}
Moreover, the constants $\alpha$ and $\beta$ follow from the expansion
\begin{equation}
\lim_{\tau_0 \to \tau_{\mathrm{max}}} P_{0 \rightarrow 1}(0,\tau_0)
= \alpha + \alpha\beta(\tau_0-\tau_{\mathrm{max}})^2 + {\mathcal{O}}(\tau_0 - \tau_{\mathrm{max}})^3
\end{equation}
and are
\begin{subequations}
\begin{align}
\label{alpha_beta}
\alpha & = \frac{a\sqrt{2\pi}}{\tau^*} \, \exp\left[ \frac{\tau^*}{a\sqrt{2\pi}} \, e^{-a^2/2} - 1 \right] \qquad \text{and} 
 \\[1ex]
\beta  & = \frac{2a^2}{(\tau^*)^2} \ln\left[\frac{a\sqrt{2\pi}}{\tau^*} \right]
\, .
\end{align}
\end{subequations}
Finally, so long as $g_1(\tau_0)$ and $g_2(\tau_0)$ satisfy Eqs.~\eqref{g_eqs}, we are free to choose their form which we do as follows
\begin{widetext}
\begin{subequations}
\begin{align}
g_1(\tau_0) &=
\Theta\left(\frac{1}{4}-\tau_0\right)
	+ \left(\frac{3}{2}-2\tau_0\right)\Theta\left(\frac{3}{4}
	- \tau_0\right)\Theta\left(\tau_0-\frac{1}{4}\right)
	\qquad \text{and}\\
g_2(\tau_0) &=
\Theta\left(\tau_0-\frac{3}{4}\right)
	+ \left(2\tau_0-\frac{1}{2}\right)\Theta\left(\frac{3}{4}
	- \tau_0\right)\Theta\left(\tau_0-\frac{1}{4}\right)
\, ,
\end{align}
\end{subequations}
\end{widetext}
where $\Theta(.)$ is the Heaviside theta function.

\begin{figure*}
\begin{center}
\includegraphics[width=1.\linewidth]{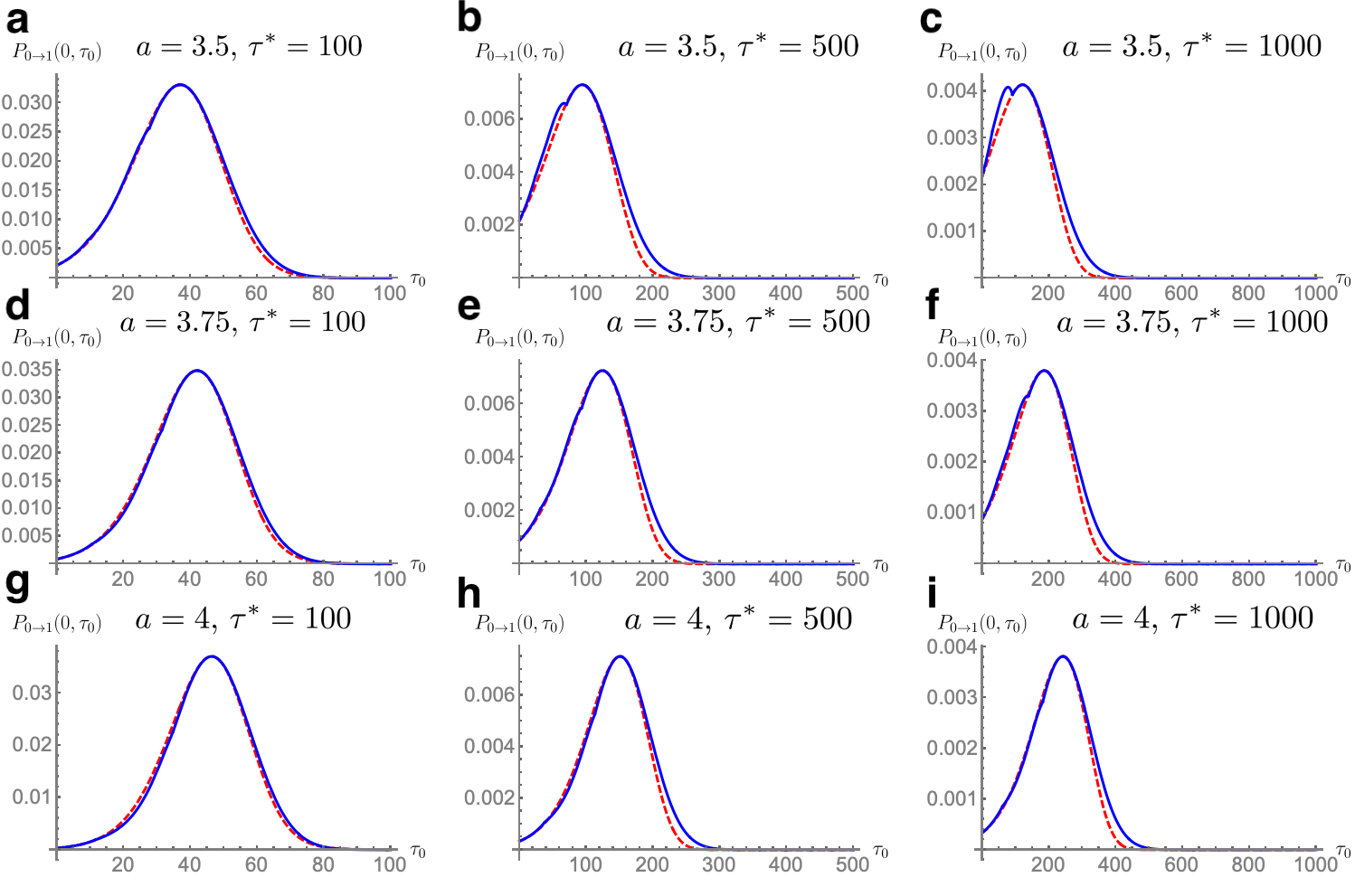}
\caption{Comparison between $P_{0 \rightarrow 1}(0,\tau_0)$ (solid blue lines) and $g(\tau_0)$ (red dashed lines) from Eqs. \eqref{eq:P01.lin} and \eqref{eq:A2} respectively. The values of $a$ and $\tau^*$ used are:
$a= 3.5$ for (a, b, c); $a= 3.75$ for (d, e, f); $a= 4$ for (g, h, i); $\tau^*= 100$ for (a, d, g); $\tau^*= 500$ for (b, e, h); and $\tau^*= 1000$ for (c, f, i).}
\label{supp_Fig1}
\end{center}
\end{figure*}

\begin{figure*}
\begin{center}
\includegraphics[width=.8
\linewidth]{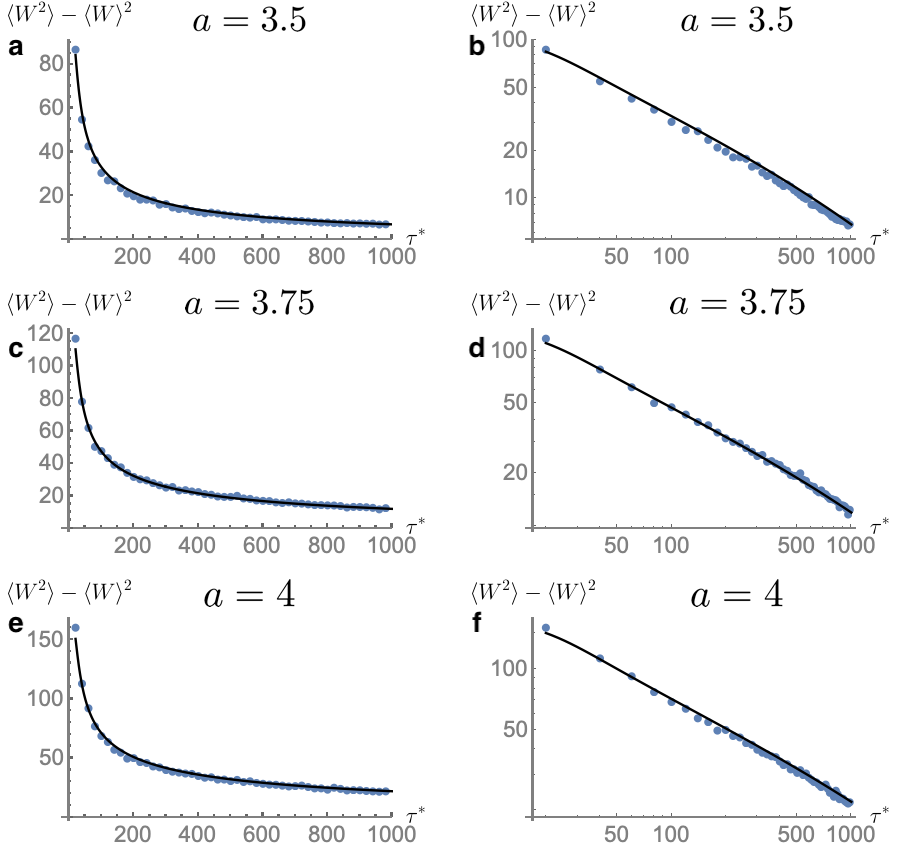}
\caption{Variance of the work obtained numerically by solving the Langevin equation \eqref{lang} (blue dots), and theoretically by replacing $P_{0 \rightarrow 1}(0,t)$ with $g(t)$ in Eq.~\eqref{app_varianceW} (black lines) \cite{note1}.  In the left (right) column we use a linear (log-log) scale.  The three values of $a$ used are: $a=3.5$ (a, b), $a=3.75$ (c, d), and $a=4$ (e, f).}
\label{supp_Fig3}
\end{center}
\end{figure*}

We approximate $P_{0 \rightarrow 1}(0,\tau_0)$ using $g(\tau_0)$, which is a sum of Gaussians multiplied by a first order polynomial and thus easily integrable upon substitution into Eqs.~\eqref{app_meanW} and \eqref{app_varianceW}.  In Fig.~\ref{supp_Fig1} we compare $P_{0 \rightarrow 1}(0,\tau_0)$ with $g(\tau_0)$, which clearly shows that the approximation provides an excellent match with the original function.  Therefore, 
as seen in Figs.~\ref{Fig6},  \ref{supp_Fig1} and \ref{supp_Fig3}, when replacing $P_{0 \rightarrow 1}(\tau_0)$ with $g(\tau_0)$ in Eqs.~\eqref{app_meanW} and \eqref{app_varianceW}), we obtain a robust analytical approximation for the mean and the variance of the work for fast erasure protocols. 


\begin{thebibliography}{}
\bibitem{note0}

In \cite{berut,berut1,jun}, the relaxation time within the potential wells is indeed much shorter than the time scale of variations in the erasure protocol. Note that the system studied in \cite{dago,dago1} is underdamped, in contrast to our model and the experiments in \cite{berut,berut1,jun}, such that the characteristic time of inertial effects plays a role too for intra-well relaxation. In their theoretical analysis the authors of \cite{dago,dago1} assumed the distribution within a potential well to be Gaussian, equivalent to a fast relaxation assumption.
\bibitem{note1}
Due to the cusp-shaped maximum between the two potential wells (see Fig.~\ref{Fig1}), the numerically integrated trajectories slightly ``overshoot'' the height of the maximum when passing from one side to the other.  The overshoot decreases as the time-discretization $\Delta t$ in our numerical integration decreases. For our choice of $\Delta t = 0.01$, we found that the ``apparent height'' of the potential barrier is about $5\%$ larger than its nominal value.   Therefore, in our theory-numerical comparisons, we calculated the theoretical curves using the corrected value for the height of the potential barrier that a particle in the numerical simulations experiences.  Thus we increased the nominal value of the height of the potential barrier $a$ by $5\%$.   Note that in its ``second role'', which is not affected by this adjustment,
the dimensionless parameter $a$ specifies the unperturbed positions of the potential wells.  We determined the correction to the barrier height independently of the simulations of the erasure process, by matching the numerically estimated survival probability of a Brownian particle in a parabolic potential of height $a$ with the analytical expression; Eq.~\eqref{eq:S01} when $F_{\textrm{erase}}(s)\equiv 0$.
We verified that the correction becomes smaller as $\Delta t$ decreases.
\bibitem{note2}
We note that many argue that the priority regarding this bound should be given to Léon Brillouin; The negentropic principle of information. {\em J. Appl. Phys. }  {\bf 24}, 1152-1163 (1953).  However, we do not opine on this matter.
\end{thebibliography}
\end{document}